\documentclass[twocolumn,showpacs,preprintnumbers,amsmath,amssymb,a4paper,superscriptaddress]{revtex4}
\usepackage{amsmath}
\usepackage{graphicx}
\usepackage{dcolumn}
\usepackage{bm}
\usepackage{booktabs}
\usepackage{calc}
\usepackage{multirow}
\usepackage{mathtools}

\DeclareMathAlphabet{\mathpzc}{OT1}{pzc}{m}{it}

\newcommand{\vek}[1]{\mathbf{#1}}
\newcommand {\beq} {\begin{eqnarray}}
\newcommand {\eeqn} [1] {\label{#1} \end{eqnarray}}

\newcommand{\rel}[1]{\rho\,}
\newcommand{\ra}[1]{\hat{\mathbf{r}}\,}
\newcommand{\ri}[1]{\mathbf{R}\,}
\newcommand{\ma}[1]{m\,}
\newcommand{\mi}[1]{M\,}
\newcommand{\xa}[1]{x}
\newcommand{\ya}[1]{y\,}
\newcommand{\za}[1]{z\,}
\newcommand{\X}[1]{X\,}
\newcommand{\Y}[1]{Y\,}
\newcommand{\Z}[1]{Z\,}

\begin{document}

\title{Improving Efficiency of Sympathetic Cooling in Atom-Ion and Atom-Atom Confined Collisions}
\date{\today}
\author{Vladimir S.Melezhik}
\email[]{melezhik@theor.jinr.ru}
\affiliation{Bogoliubov Laboratory of Theoretical Physics, Joint Institute for Nuclear Research, Dubna, Moscow Region 141980, Russian Federation}
\affiliation{Dubna State University, 19 Universitetskaya Street, Dubna, Moscow Region 141982, Russian Federation}

\date{\today}
\begin{abstract}\label{txt:abstract}
We propose a new way for sympathetic cooling of ions in an electromagnetic Paul trap: it implies the use for this purpose of cold buffer atoms in the region of atom-ion confinement-induced resonance (CIR). The problem is that the unavoidable micromotion of the ion and the long-range nature of its interaction with the environment of colder atoms in a hybrid atomic-ion trap prevent its sympathetic cooling. We show that the destructive effect of ion micromotion on its sympathetic cooling can however be suppressed in the vicinity of the atom-ion CIR. The origin of this is the "fermionization" of the atom-ion wave function near CIR, where the atom-ion pair behaves as a pair of noninteracting identical fermions. This prevents the complete approach of the atom with the ion near resonance and does not enhance the ion micromotion, which interferes with its sympathetic cooling. We investigate the effect of sympathetic cooling around CIRs in atom-ion and atom-atom confined collisions within the qusiclassical-quantum approach using the Li-Yb$^+$ and Li-Yb confined systems as an example. In this approach, the Schr\"odinger equation for a cold light atom is integrated simultaneously with the classical Hamilton equations for a hotter heavy ion or atom during collision. We have found the region near the atom-ion CIR where the sympathetic cooling of the ion by cold atoms is possible in a hybrid atom-ion trap. We also show that it is possible to improve the efficiency of sympathetic cooling in atomic traps by using atomic CIRs.
\end{abstract}

\maketitle

\section{INTRODUCTION}

In the last decade there has been great interest in ultracold hybrid atom-ion systems, which is caused by new opportunities that arise here for control and  simulation of various quantum processes and phenomena: formation of novel molecular states\cite{Cote, Shuher}, simulation of electron-phonon coupling in solid state physics~\cite{Bissbort}, Feshbach resonances~\cite{Idziaszek, Moszynski}, quantum information processing~\cite{Doerk, Secker}  etc. \cite{Tomza}. However, a realization of the hot proposals with cold atoms and ions~\cite{Tomza} is impeded by the unremovable ion micromotion caused by the time-dependent radio frequency (RF) fields of the Paul traps used for confining ions in the hybrid atom-ion systems \cite{Grier, Meir, Vuletic, Furst, Tomza}. Particularly, it is known that the micromotion of the ion and the long-range nature of its interaction with the environment of colder atoms in a hybrid atomic-ion trap prevent the desired effect of sympathetic cooling of ions~\cite{Vuletic, Tomza}. Despite the successes achieved in sympathetic cooling of ions in hybrid atomic-ion systems in the millikelvin range and above~\cite{Zipkes, Ravi, Harster, Haze,Smith, Tomza}, as well as the proposed promising schemes for cooling to lower energies~\cite{Secker, Furst, Vuletic, Feldker, Kleinbach, Prudnikov}, the problem of sympathetic cooling to lower energies in these systems is still pending.

In this paper, we propose a new way for sympathetic cooling of ions in an electromagnetic Paul trap: to apply for this purpose  buffer cold atoms in the region of the atom-ion confinement-induced resonance (CIR). We show that the negative effect of micromotion on sympathetic ion cooling can be suppressed in the vicinity of the atom-ion CIR. Atom-ion CIRs were predicted in ~\cite{MelNegr} and the influence of ion micromotion on the CIR position in Li-Yb$^+$ was investigated in the subsequent paper~\cite{Melezhik2019}. It was shown that the CIR occurs when the ratio of the transverse width of the atomic trap $a_{\perp}$ and the s-wave atom-ion scattering length in free space $a_s$ coincides with the value  $a_{\perp}/a_s=1.46$. Earlier, this condition was predicted~\cite{Olshanii} and subsequently confirmed in experiment~\cite{Haller2010} for atomic Cs waveguide-like traps. To describe the dynamics of a quantum particle near CIR, the 1D Fermi quasipotential with an effective coupling constant $g_{1D}(a_{\perp}/a_s)$ proposed in ~\cite{Olshanii} is successfully used. Atomic CIRs~\cite{Olshanii, Bergeman, Moore, Kim, Naidon, Saeidian, Mel2011, Giannakeas} aroused great interest and stimulated research in this direction due to the possibility of using such resonances to tune effective interatomic interactions in a wide range - from super strong attraction $g_{1D}\rightarrow -\infty$ to super strong repulsion $g_{1D}\rightarrow +\infty$~\cite{Haller2010, Gunter, Frolich, Kinoshita, Peredes, Haller2009, Selim}. It is also known that at the point $a_{\perp}/a_s$ of CIR the divergence of coupling constant $g_{1D}(a_{\perp}/a_s)$ (and the total reflection) leads to "fermionization" of the relative wave-function of the colliding pair whose square modulus behaves the same as for two noninteracting identical fermions~\cite{Girardeau, Olshanii, Kinoshita, Peredes, Selim}. This can lead to some compensation of the long-range character of the atom-ion interaction and, as a consequence, to suppression of the micromotion-induced heating during collisions confined by the atom-ion trap.  Here, we investigate how the "fermionization" can "truncate" the effective atom-ion interaction and the possibility of using this effect for improving the sympathetic cooling of the ions by buffer cold atoms in hybrid atom-ion traps.
We investigate the effect of sympathetic cooling around CIRs in atom-ion confined collisions within the qusiclassical-quantum approach~\cite{MelSchm, Melezhik2001, MelezhikCohen, MelSev, Melezhik2019} using the $^6$Li-$^{171}$Yb$^+$ pair in the hybrid atom-ion trap as an example, which is currently under intense experimental investigations~\cite{JogerPRA17,FuertsPRA18,Feldker}. It is assumed that this specific atom-ion pair is most perspective for sympathetic cooling and reaching the s-wave regime of ions in Paul traps~\cite{Vuletic, Feldker}. The following problem is considered: a hot ion confined in a time-dependent RF Paul trap with linear geometry collides with the cold atom constrained to move into a quasi-one-dimensional waveguide within the ion trap (see Fig.~\ref{fig:sketch}). In our approach~\cite{Melezhik2019, Mel2019}, the Schr\"odinger equation for a cold light atom is integrated simultaneously with the classical Hamilton equations for hotter heavy ion during collision. We have found the regions near the atom-ion CIR where the sympathetic cooling of the ion by cold atoms is possible in a hybrid atom-ion trap.  The possibility of sympathetic cooling of a heavy hot atom by light cold atoms near atomic CIR is also investigated for the case of a Li-Yb mixture confined by an atomic trap. We show that it is possible to improve the efficiency of sympathetic cooling in atomic traps by using atomic CIRs.

In the next section, our theoretical approach and the principal elements of the computational scheme are discussed. In Section III, the results and discussions are presented. The concluding remarks are given in the last section. Some technical details of the computations are discussed in the Appendix.

\section{PROBLEM FORMULATION AND COMPUTATIONAL SCHEME}
\begin{figure}
\centering\includegraphics[scale=0.35]{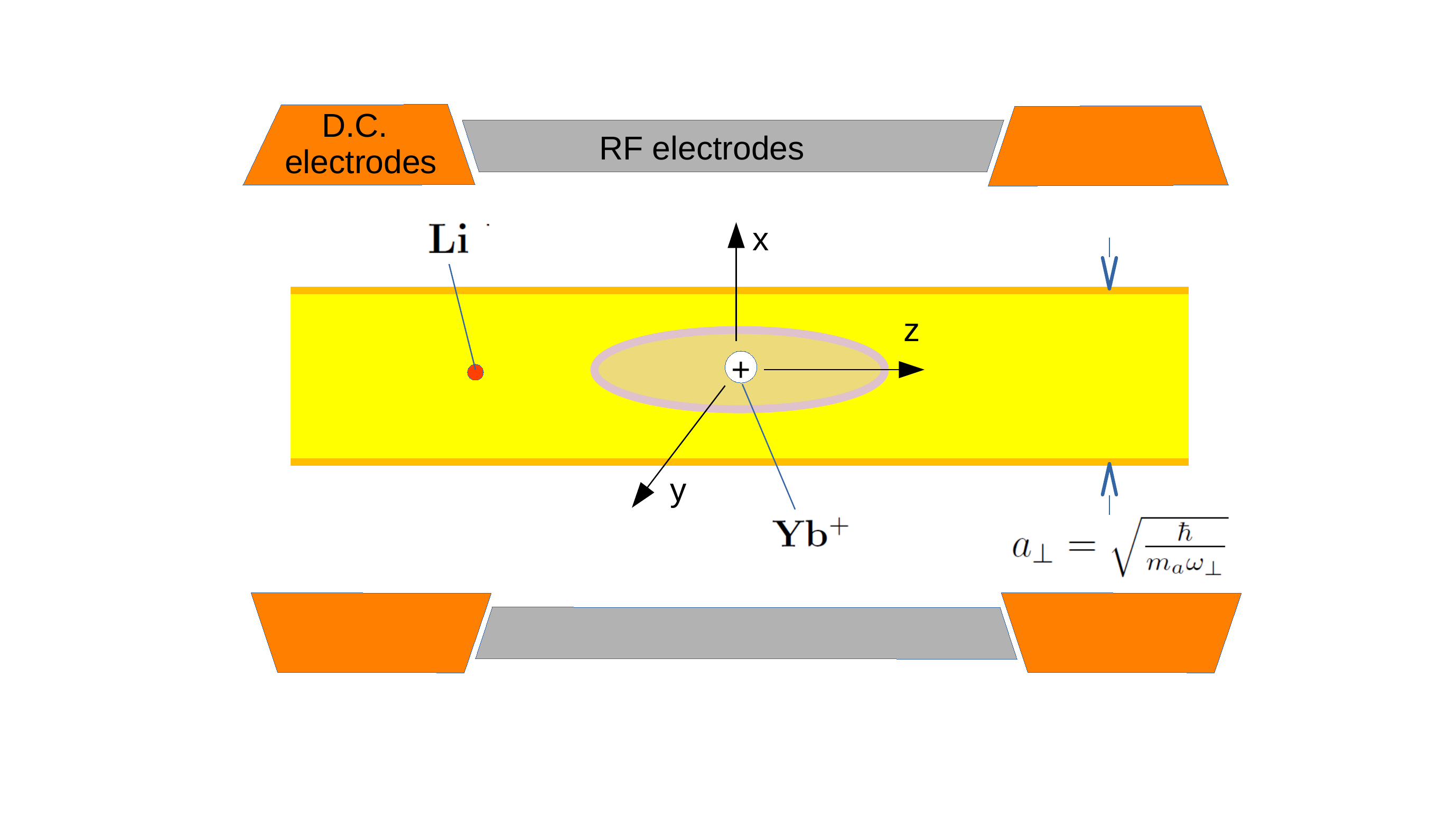}
\caption{(color online) Schematic representation of the atom-ion system confined in a hybrid trap. Here, the ion  is situated in the cloud of cold atoms confined by an optical atomic trap inside the electromagnetic Paul trap. The time-dependent RF field confines the ion transversally, whereas longitudinally a static confinement is formed by the DC field. The dimensions of the confinement region of the ion are determined by the frequencies of the Paul trap $\omega_i$ and $\Omega_{rf}$. The atomic waveguide along the longitudinal axis, $z$, of the linear Paul trap confines the atoms in the transverse $x,y$ directions. The width of the atomic trap $a_{\perp}=\sqrt{\hbar/(m_a\omega_{\perp})}$ is determined by the frequency $\omega_{\perp}$ of the harmonic approximation for the trap shape. Inside the hybrid trap occur paired atom-ion collisions.}
\label{fig:sketch}
\end{figure}

A schematic view of the system under investigation is given in Fig.~\ref{fig:sketch}. A Li atom of  cold atomic cloud confined in the transverse direction by a harmonic potential of an optical trap
\begin{align}
\label{eq:Uatom}
V(\vek{r}_a) = \frac{m_a\omega_{\perp}}{2}\left(x_a^2+y_a^2\right)
\end{align}
collides with the Yb$^+$ ion in the potential created by the linear RF Paul trap~\cite{LeibfriedRMP03,Feldker, JogerPRA17, FuertsPRA18}
\begin{align}
\label{eq:Uion}
U(\vek{r}_i,t) = \frac{m_i\omega_i^2}{2}\left(z_i^2-\frac{x_i^2+y_i^2}{2}\right) \nonumber\\
+\frac{m_i\Omega_{rf}^2}{2} q \cos(\Omega_{rf} t)\left(\frac{y_i^2}{2}-\frac{x_i^2}{2}\right)\,.
\end{align}
The interaction potentials (\ref{eq:Uatom}),(\ref{eq:Uion}) of an atom and an ion with a hybrid trap depend on the transverse frequency of the atomic trap $\omega_{\perp}$, which determines its transverse size $a_{\perp} = \sqrt{\hbar/(m_a\omega_{\perp})}$, and the frequencies of the Paul trap $\omega_i$, $\Omega_{rf}$. Here, $\omega_i = \Omega_{rf}\sqrt{a/2}$ is the so-called secular frequency~\cite{LeibfriedRMP03}, 
$q$ and $a$ are dimensionless geometric parameters (i.e. $a=0.002\ll q^2=0.08^2 <1$), which in our calculations were chosen according to the installation parameters~\cite{JogerPRA17,FuertsPRA18,Feldker}. The last term in the potential (\ref{eq:Uion}), which depends on the frequency $\Omega_{rf}$, causes RF oscillations of the ion, i.e. sets its micromotion. The vectors $\vek {r}_a$ and $\vek{r}_i $ set the coordinates of the atom and the ion, and $m_a$ and $m_i$ are the masses of the atom and the ion, respectively. We assume that the axis of the waveguide in which, the colliding atom is travelling,  is precisely the $z$-axis of the Paul trap (see Fig.~\ref{fig:sketch}). The origin is at the center of the Paul trap.

In our work~\cite{Melezhik2019}, the quantum-quaiclassical approach~\cite{MelSchm, Melezhik2001, MelezhikCohen, MelSev} was extended and adapted for quantitative description of pair collisions of light slow Li atoms with heavy Yb$^+$ ions in the confined geometry of the hybrid atom-ion trap defined by  potentials (\ref{eq:Uatom}),(\ref{eq:Uion}).  In this approach, the problem is reduced to the simultaneous integration of a system of coupled quantum and classical equations: the time-dependent Schr\"odinger equation, that describes the collisional dynamics of an atom confined in an optical trap (\ref{eq:Uatom}) with an ion, and the classical Hamilton equations, describing the vibrations of an ion in a Paul trap (\ref{eq:Uion})and its perturbation during collision with an atom.

The time-dependent Schr\"odinger equation for the wave function of the atom $\psi(\vek{r}_a,t)$ is written in the form
\begin{align}
\label{eq:Schr}
i\hbar \frac{\partial}{\partial t}\psi(\vek{r}_a,t) = [-\frac{\hbar^2}{2m_a}\triangle_a + V(\vek{r}_a) \nonumber\\
+ V_{ai}(\mid \vek{r}_a-\vek{r}_i(t)\mid)]\psi(\vek{r}_a, t) \,,
\end{align}
where the potential $V_{ai}$ defines the atom-ion interaction.
The classical Hamiltonian describing an ion in a Paul trap is given by
\begin{align}
H_i^{trap}(\vek{p}_i,\vek{r}_i,t) = \frac{\vek{p}_i^2}{2 m_i} + U(\vek{r}_i,t)\,,
\end{align}
where $\vek{p}_i(t)$ is the ion momentum.
When the atom is confined in the optical waveguide within the Paul trap, the ion experiences its presence via the atom-ion interaction $V_{ai}(\vert\vek{r}_a-\vek{r}_i(t)\vert)$ during the collision. Therefore, the full classical ion Hamiltonian is given by
\begin{align}
\label{eq:Ham}
H_i(\vek{p}_i,\vek{r}_i,t;\vek{r}_a) = H_i^{trap}(\vek{p}_i,\vek{r}_i,t)
+ \langle V_{ai}(\vert\vek{r}_a-\vek{r}_i(t)\vert)\rangle \,,
\end{align}
where
$$
	\langle V_{ai}(\vert\vek{r}_a-\vek{r}_i(t)\vert)\rangle = \langle\psi(\vek{r}_a,t)\vert V_{ai}(\vert\vek{r}_a-\vek{r}_i(t)\vert)\vert\psi(\vek{r}_a,t)\rangle
$$
is the quantum mechanical average of the atom-ion interaction over the atomic density instantaneous distribution $|\psi(\vek{r}_a,t)|^2$. Thus, the ion Hamiltonian (\ref{eq:Ham}) defined in such a way has a parametric dependence on the atom position $\vek{r}_a(t)$. It leads at the moment of the atom-ion collision to the strong non-separability of the Hamilton equations
\begin{align}
\label{eq:Hamilton}
\frac{d}{d t}\vek{p}_i & = -\frac{\partial}{\partial \vek{r}_i}H_i (\vek{p}_i,\vek{r}_i,t;\vek{r}_a)\nonumber\\
\frac{d}{d t}\vek{r}_i & = \frac{\partial}{\partial \vek{p}_i}H_i (\vek{p}_i,\vek{r}_i,t;\vek{r}_a)
\end{align}
describing the ion dynamics and its strong coupling with the time-dependent Schr\"odinger equation (\ref{eq:Schr}). As a consequence, it requires sufficient stability of the computational scheme~(see Appendix).

The set of classical equations (\ref{eq:Hamilton})for the ion variables $\vek{r}_i$ together with the Schr\"odinger equation (\ref{eq:Schr}) for the atomic wave function $\psi(\vek{r}_a,\vek{r}_i(t))$ form a complete set of dynamical equations for describing the atom-ion collision dynamics in a hybrid confining trap~\cite{Melezhik2019}. In the present study we consider collisions of a light atom with a much heavier ion in the range of very low atomic colliding energies $E_{\mathrm{coll}}$
(ultracold atoms), where the relation $p_a=\sqrt{2m_a E_{\mathrm{coll}}} \ll p_i$ for their momentums is satisfied.
In addition,  we require that $E_i = p_i^2/(2 m_i) \gg \hbar \omega_i$, which further justifies the application of the classical description for the ion.

At the instant of collision
($ r = |\vek{r}_a- \vek{r}_i(t)| \rightarrow 0 $), the equations of the system (\ref{eq:Schr}, \ref{eq:Hamilton}) are strongly coupled by the potential of the atomic-ion interaction, which in our scheme is chosen in the regularized form~\cite{Krych2015,Melezhik2019}
\begin{align}
\label{eq:vai}
V_{ai}(r) = - \frac {(r^2-c^2)}{(r^2 + c^2)}\frac {C_4}{(r^2 + b^2)^2 }\,\,,
\end{align}
and decay into independent equations for an atom and an ion in the asymptotic region before and after the collision, where $r \rightarrow \infty $ and $V_{ai}$ - interaction disappears:
$V_{ai}(r) \rightarrow -C_4/r^4 \rightarrow 0 $. The dispersion coefficient $C_4$ is a known parameter for the Li-Yb$^+$ pair, and by varying the free parameters $b$ and $c$ we change the intensity of the atomic-ion interaction, i.e. define the s-wave atomic-ion scattering length in free space $a_s$, which can be experimentally tuned. Thus, for tuning the interatomic interactions (the scattering length $a_s$) in atomic traps the magnetic Feshbach resonances are successfully used~\cite{Chin2010,Haller2010,Selim}. The prospects for their use in hybrid atomic-ion systems are also discussed~\cite{Moszynski,Feldker}.

To integrate simultaneously equations (\ref{eq:Schr}, \ref{eq:Hamilton}), we need proper initial conditions with physical significance. At the beginning of the collisional process, the atom and the ion are assumed to be far away from each other so that they do not interact ($V_{ai}=0$). In particular, the atom is initially in the ground state of the atomic trap with the longitudinal colliding energy, that is, $E_{\mathrm{coll}} \ll 2\hbar\omega_{\perp}$
\begin{align}
\label{eq:atom0}
\psi(\vek{r}_a,t=0)= N\phi_0(\rho_a)e^{-\frac{(z_a-z_0)}{2a_z}}e^{ikz_a}\,,
\end{align}
whereas the ion performs fast (with respect of atom motion) oscillations in the Paul trap with mean transversal $\langle E_{\perp}\rangle=\langle E_{ix}\rangle +\langle E_{iy}\rangle $ and longitudinal $\langle E_{\parallel}\rangle =\langle E_{iz}\rangle$ energies, which can be fixed by the  proper choice of the ion initial conditions
\begin{align}
\label{eq:ion0}
\vek{r}_i(t=0) &= \vek{r}_0 \nonumber\\
p_{i,x}(t=0) &= \sqrt{2 m_i E_{ix}^{(0)}},\nonumber\\
p_{i,y}(t=0) &= \sqrt{2 m_i E_{iy}^{(0)}},\nonumber\\
p_{i,z}(t=0) &= \sqrt{2 m_i E_{iz}^{(0)}}\,.
\end{align}
In Eq.(\ref{eq:atom0}) $\phi(\rho_a)$ is the wave function of the ground state of a two-dimensional harmonic oscillator approximating an atomic trap potential and N is the normalization coefficient. The parameter $a_z$ specifies the width of the initial wave packet of the atom (\ref{eq:atom0}) in the longitudinal direction, which must be wide enough for the wave packet to be sufficiently monochromatic and its spreading in time could be neglected~\cite{Melezhik2019}. The initial position of the wave packet (\ref{eq:atom0}) is set by the value of $z_0$ so that at the initial moment $t = 0$ the atomic-ion interaction $V_{ai}$ can be neglected.

The initial conditions (\ref{eq:atom0},\ref{eq:ion0})  set the initial state of a noninteracting atom-ion system: an ion performing a finite motion in a Paul trap with given mean energies $\langle E_{\parallel}\rangle$ and $\langle E_{\perp}\rangle$ and a slow atom in the ground state  $\phi_0(\rho_a)$, which moves in the longitudinal direction of the optical trap with a velocity $v_a=\hbar k/m_a=\sqrt{2E_{coll}/m_a}$.
Since the atom approaches the region of interaction with the ion very slowly ($E_{\mathrm{coll}} /\hbar \ll \omega_{\perp} \ll \omega_i,\Omega_{rf}$), the initial position of the ion does not influence the scattering process itself, which depends only on $\langle E_{\perp}\rangle$ and $\langle E_{\parallel}\rangle$.

As a result of the integration of the system of equations (\ref{eq:Schr}, \ref{eq:Hamilton}), the wave packet $\psi(\vek{r}_a,t)$ is calculated. Asymptotically it has the following behavior at $t\rightarrow+\infty$
\begin{align}
\label{eq:atomout}
\psi(\vek{r}_a,t)
&\mathop{\longrightarrow}\limits_{z_a \rightarrow
+\infty}\ (1+f^+) N \phi_0(\rho_a)\chi(z_a-z_0)e^{-ik_fz_a}
\end{align}
in the asymptotic region $r_a\rightarrow +\infty$, where $f^+(a_{\perp}/a_s)$ is the forward scattering amplitude describing the atom-ion collision confined by the hybrid trap. The longitudinal part $\chi(z_a-z_0)$ of the atomic wave packet describes the atom motion in  $z$-direction, namely the spreading of the initial Gaussian wave packet $\exp[-\frac{(z_a-z_0)}{2a_z}]$ (\ref{eq:atom0}). Due to the choice of the longitudinal width $a_z$ of the initial wave packet rather large we achieve its sufficient monochromaticity along z-direction $k\simeq k_f$. This provides insignificant deformation of the envelope $\chi(z_a-z_0)$ in (\ref{eq:atomout}) with respect of its initial form in (\ref{eq:atom0}), what permits to calculate with enough accuracy the scattering amplitude~\cite{Melezhik2019}
\begin{align}
\label{eq:overlap}
\langle\psi^{(0)+}(t)\vert\psi(t)\rangle\mathop{\longrightarrow}\limits_{t
\rightarrow +\infty}1+f^{+}(k)\,.
\end{align}
Here, the wave packet  $\psi^{(0)+}(t)$ is calculated by independent integration of the Schr\"odinger equation (\ref{eq:Schr}) with the same initial conditions (\ref{eq:atom0}) and $V_{ai}=0$~\cite{Melezhik2019}, which has during the special choice of $a_z$ mentioned above the following asymptotic behavior
\begin{align}
\label{eq:ansatz-withoution}
\psi^{(0)+}(\vek{r}_a,t\rightarrow +\infty) = N \varphi_{0}(\rho_a)\chi(z_a-z_0))\,e^{ikz_a}\,.
\end{align}

The scattering amplitude determines the quasi-1D atomic-ion coupling constant by~\cite{Olshanii,Bergeman}
\begin{align}
\label{eq:g1D}
g_{1D}=\mathop{\lim}\limits_{k
\rightarrow 0}
\frac{\hbar^2 k}{m_a}\frac{Re[f^{+}(k)]}{Im[f^{+}(k)]}\,.
\end{align}
The constant $g_{1D}$ is the most relevant parameter for analysing confined scattering close to a CIR, where $g_{1D}\rightarrow \pm\infty$~\cite{Olshanii,Bergeman,Kim,Saeidian}.
 and the transmission
\begin{align}
\label{eq:T}
T(a_{\perp}/a_s) = \mid 1+f^+(a_{\perp}/a_s)\mid^2 \rightarrow 0 \,\,.
\end{align}

The trajectory of the ion $\vek{r}_i(t)$, its momentum $\vek{p}_i(t)$ and kinetic energy $E_i(t)=\vek{p}_i^2(t)/(2m_i)$ are also calculated, which enter stable trajectories after colliding with an atom in the asymptotic region, if the ion does not leave the Paul trap after the collision.

\section{RESULTS and DISCUSSION}
\subsection{Stability of ion motion near CIR in hybrid trap}

To test the assumption about the possibility of suppressing micromotion-induced heating of an ion near the "fermionization" point $a_{\perp}/a_s=1.46$, we first have considered the collision of a cold $^6$Li atom with an $^{171}$Yb ion at rest in the center of the hybrid trap (see Fig.1), i.e. with $\vek{r}_0=E_i^{(0)}=0$ in the initial conditions for ion (\ref{eq:ion0}), for different parameters of the effective atomic-ion interaction by varying the ratio $a_{\perp}/a_s$.  The ratio $a_{\perp}/a_s$ was varied at a fixed width of atomic trap $a_{\perp}=\sqrt{\hbar/(m_a \omega_{\perp})}$ by changing the value of the atom-ion scattering length $a_s$. This was achieved by varying the parameters $b$ and $c$ in the atom-ion interaction potential (\ref{eq:vai}) at the interval $0.1R^{*2}\leq b^2 \leq 0.3R^{*2}$ and $c^2\sim 0.1 R^{*2}$.
Hereafter, we  use the units of the problem: $R^*=\sqrt{2\mu C_4}/\hbar$, $E^*=\hbar^2/(2\mu R^{*2})$, $p^*=\sqrt{2\mu E^*}$, $\omega^*=E^*/\hbar$ and $t_{\perp}=2\pi/\omega_{\perp}$, where $\mu$ is atom-ion reduced mass. The calculations were performed for the atom initially in the ground state with transversal $E_{\perp}=\hbar\omega_{\perp}=0.02E^*$ and longitudinal colliding energy $E_{coll}/k_B=0.004E^*/k_B=11$nK for $\omega_{\perp}=0.02\omega^*=2\pi\times 7.1$kHz. The Paul trap frequencies were chosen as $\Omega_{rf}=2\pi\times 2$MHz and $\omega_i=2\pi\times 63$kHz ~\cite{JogerPRA17,FuertsPRA18,Feldker}.

\begin{figure*}
\parbox{0.95\textwidth}{
\includegraphics[width=0.3\textwidth,height=0.3\textwidth]{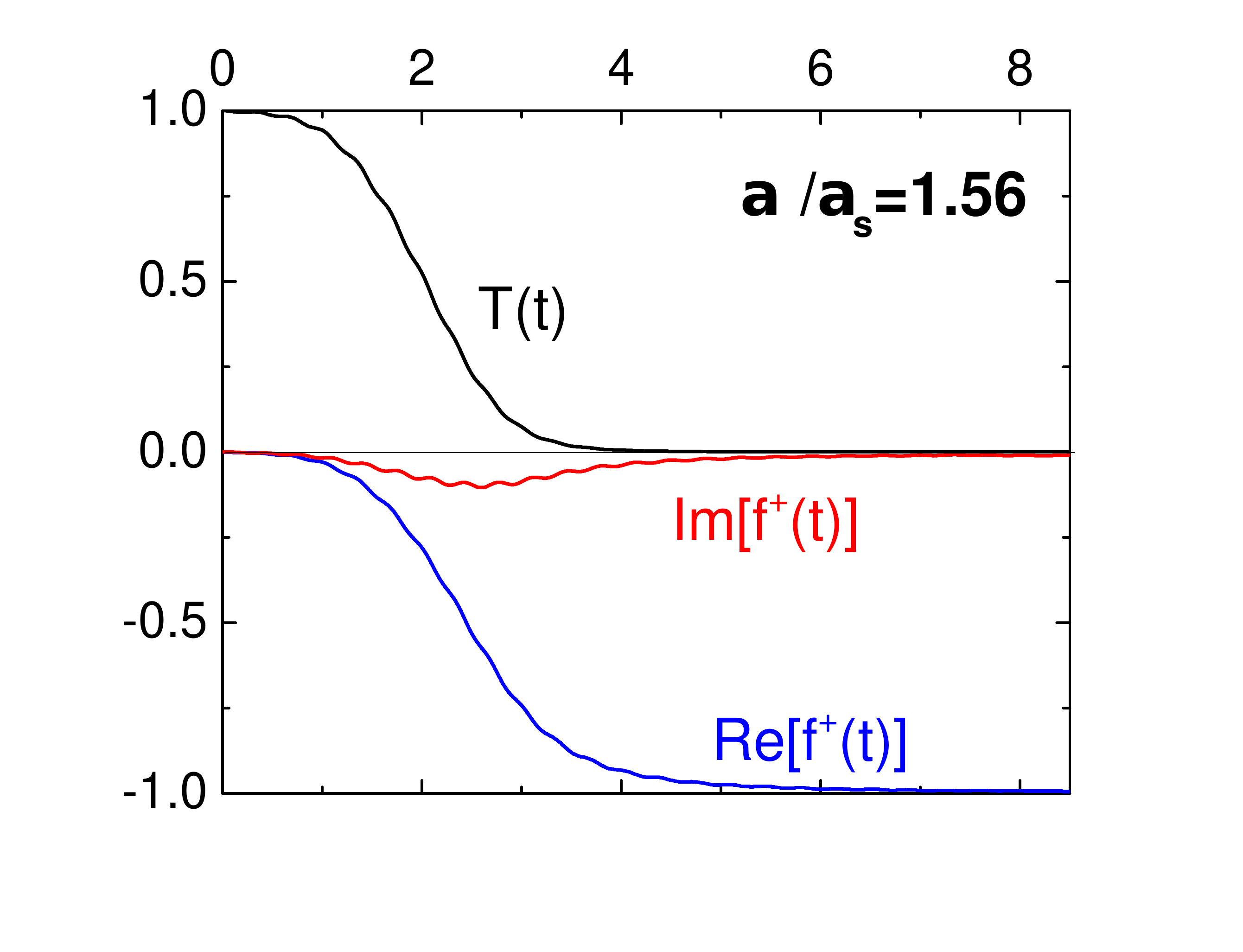}
\includegraphics[width=0.3\textwidth,height=0.3\textwidth]{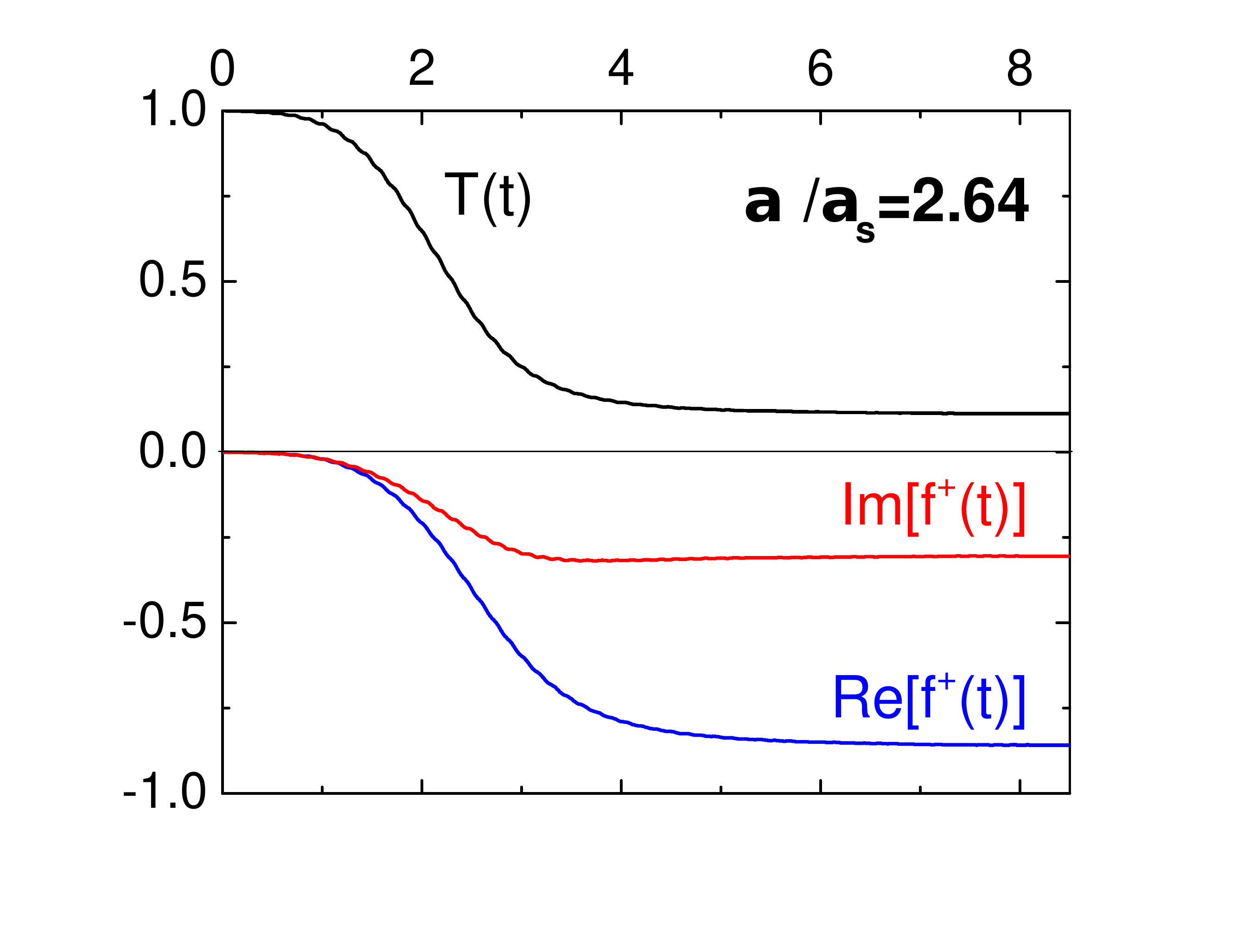}
\includegraphics[width=0.3\textwidth,height=0.3\textwidth]{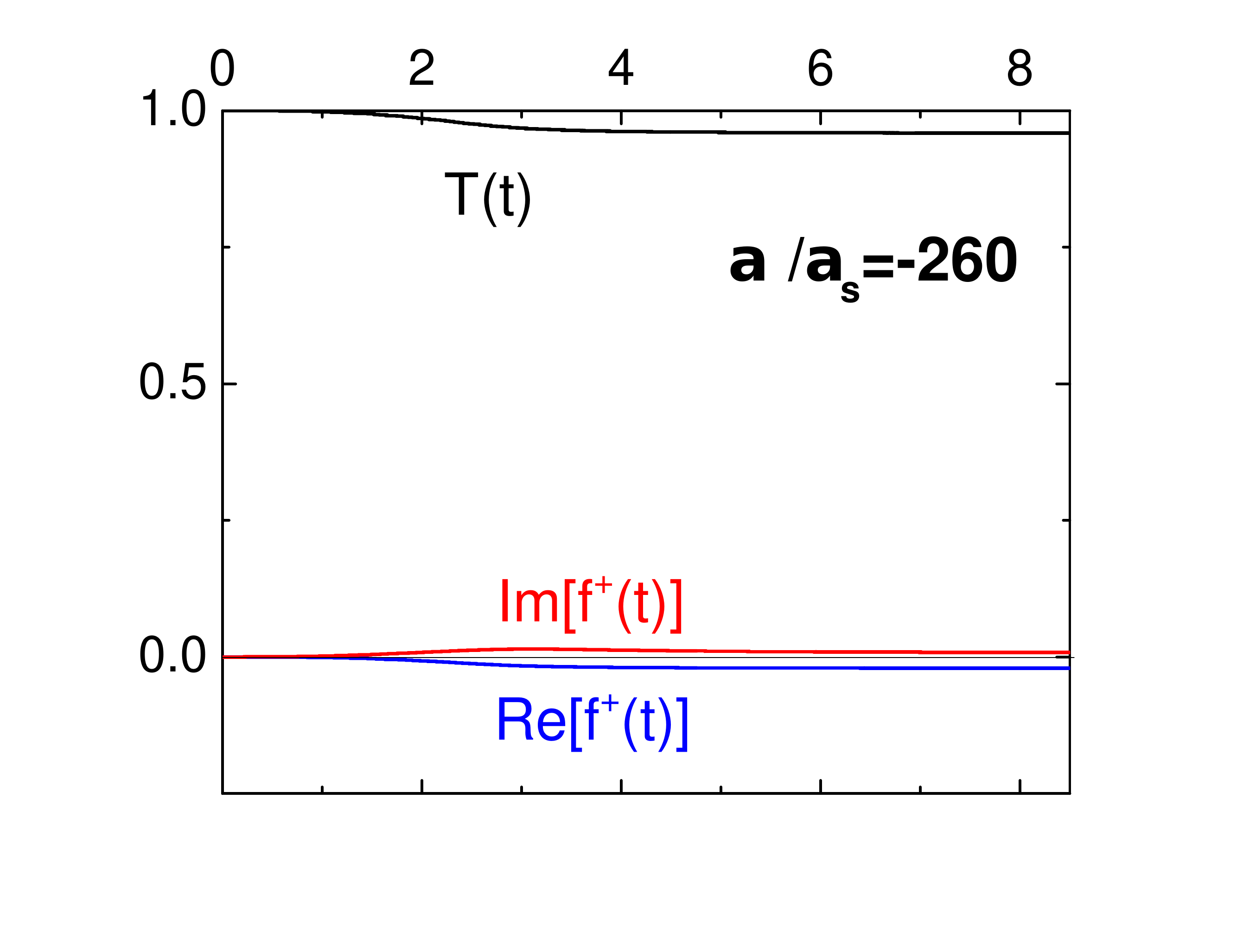}
\vspace{-1.2cm}}
\parbox{0.95\textwidth}{
\includegraphics[width=0.3\textwidth,height=0.3\textwidth]{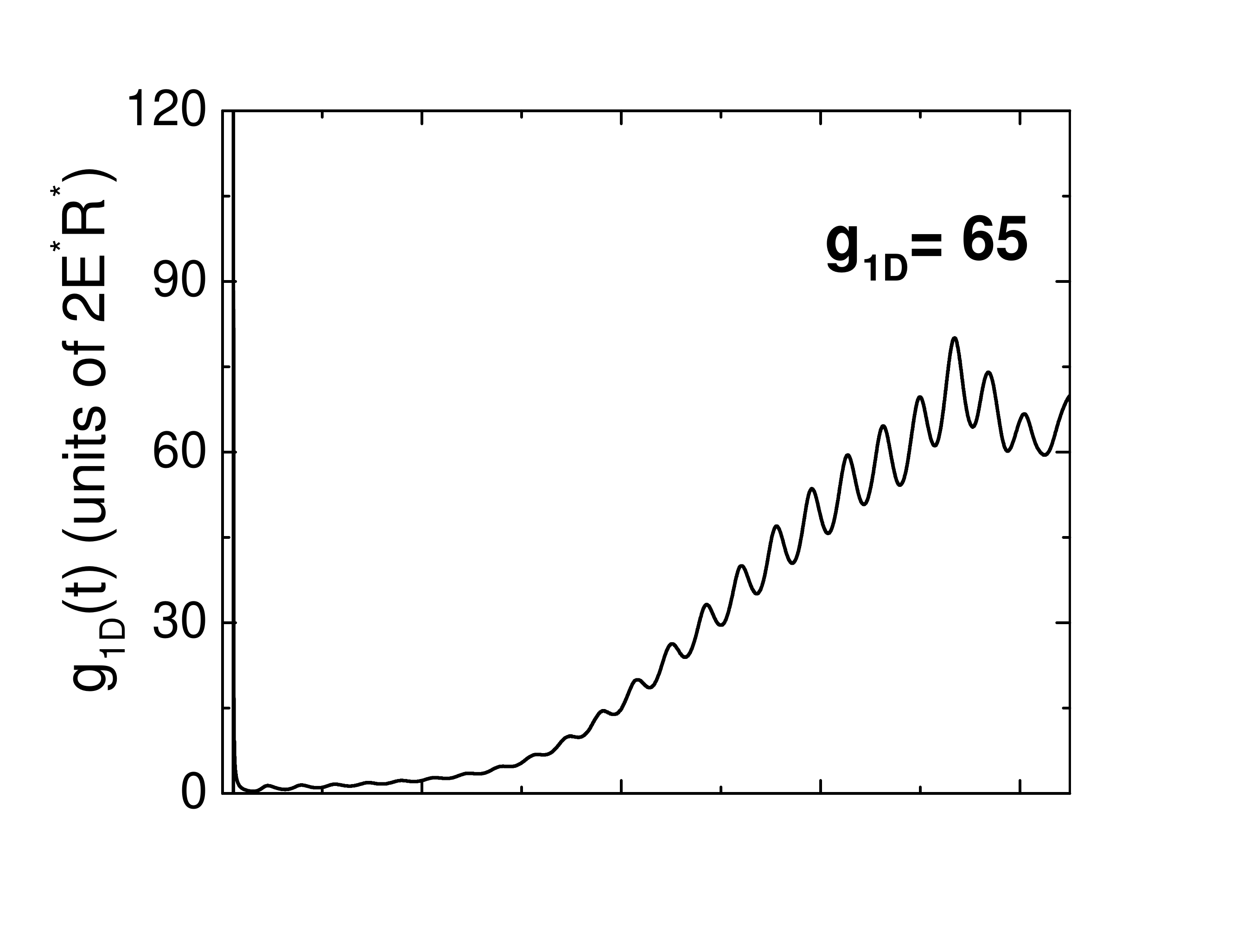}
\includegraphics[width=0.3\textwidth,height=0.3\textwidth]{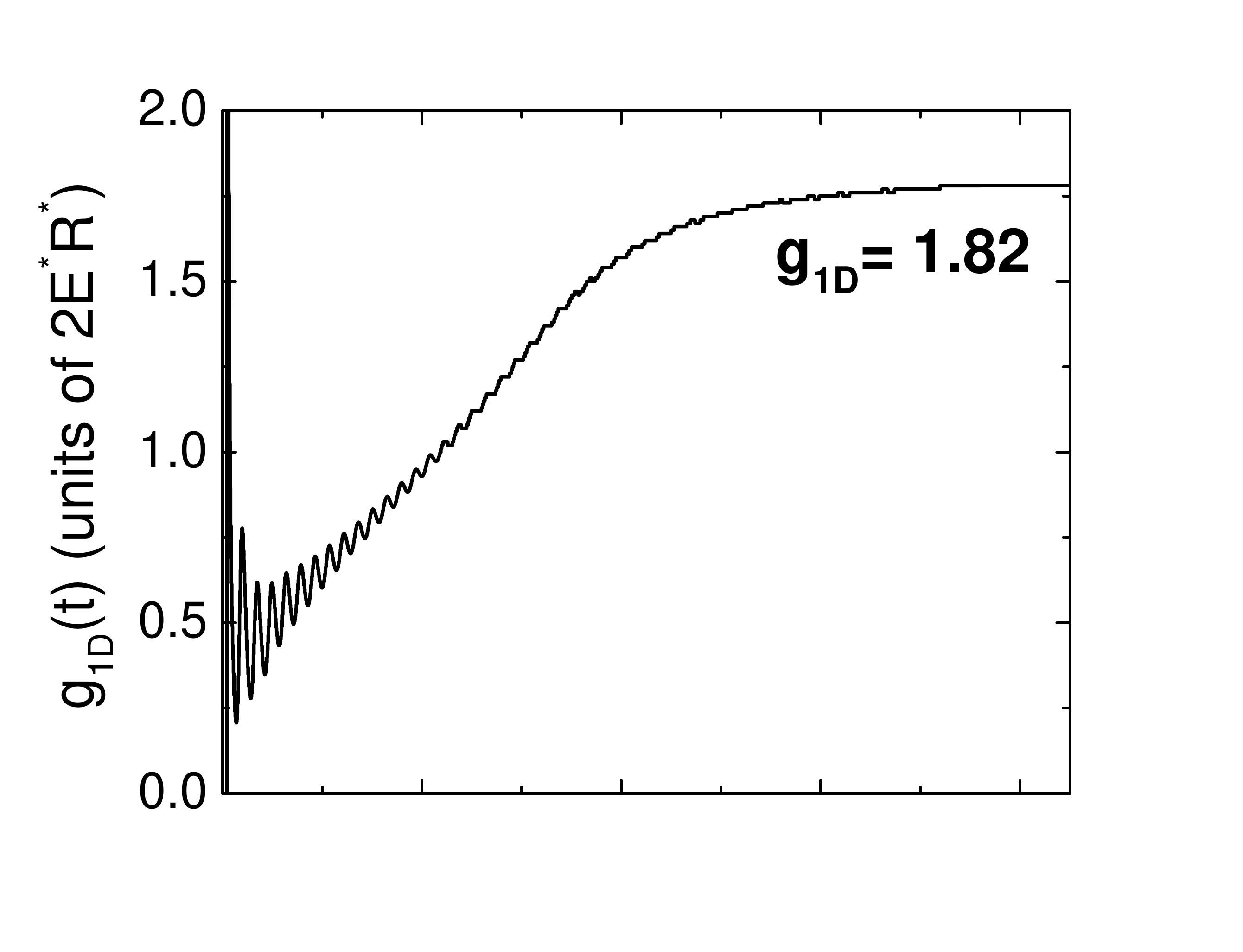}
\includegraphics[width=0.3\textwidth,height=0.3\textwidth]{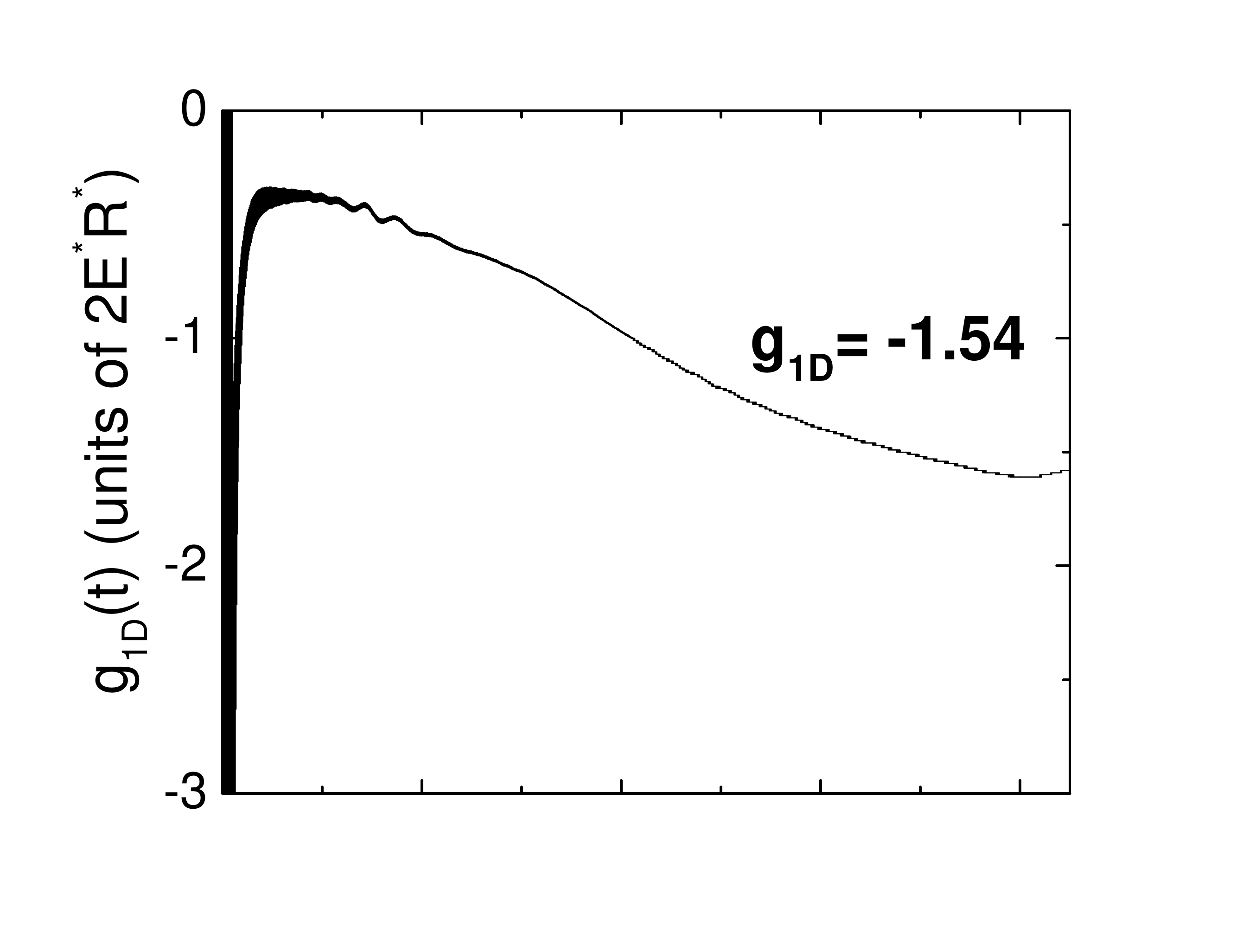}
\vspace{-1.2cm}}
\parbox{0.95\textwidth}{
\includegraphics[width=0.3\textwidth,height=0.3\textwidth]{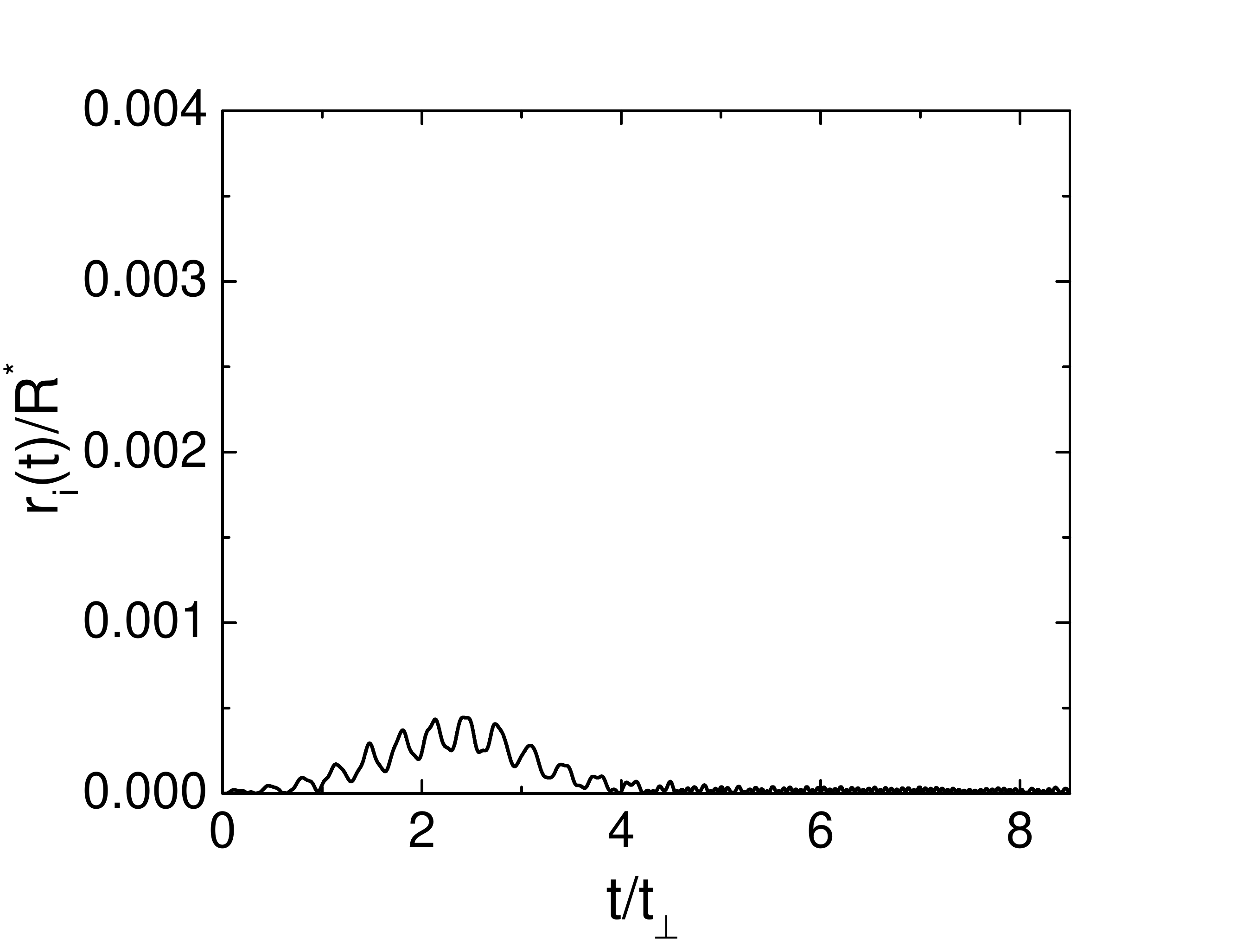}
\includegraphics[width=0.3\textwidth,height=0.3\textwidth]{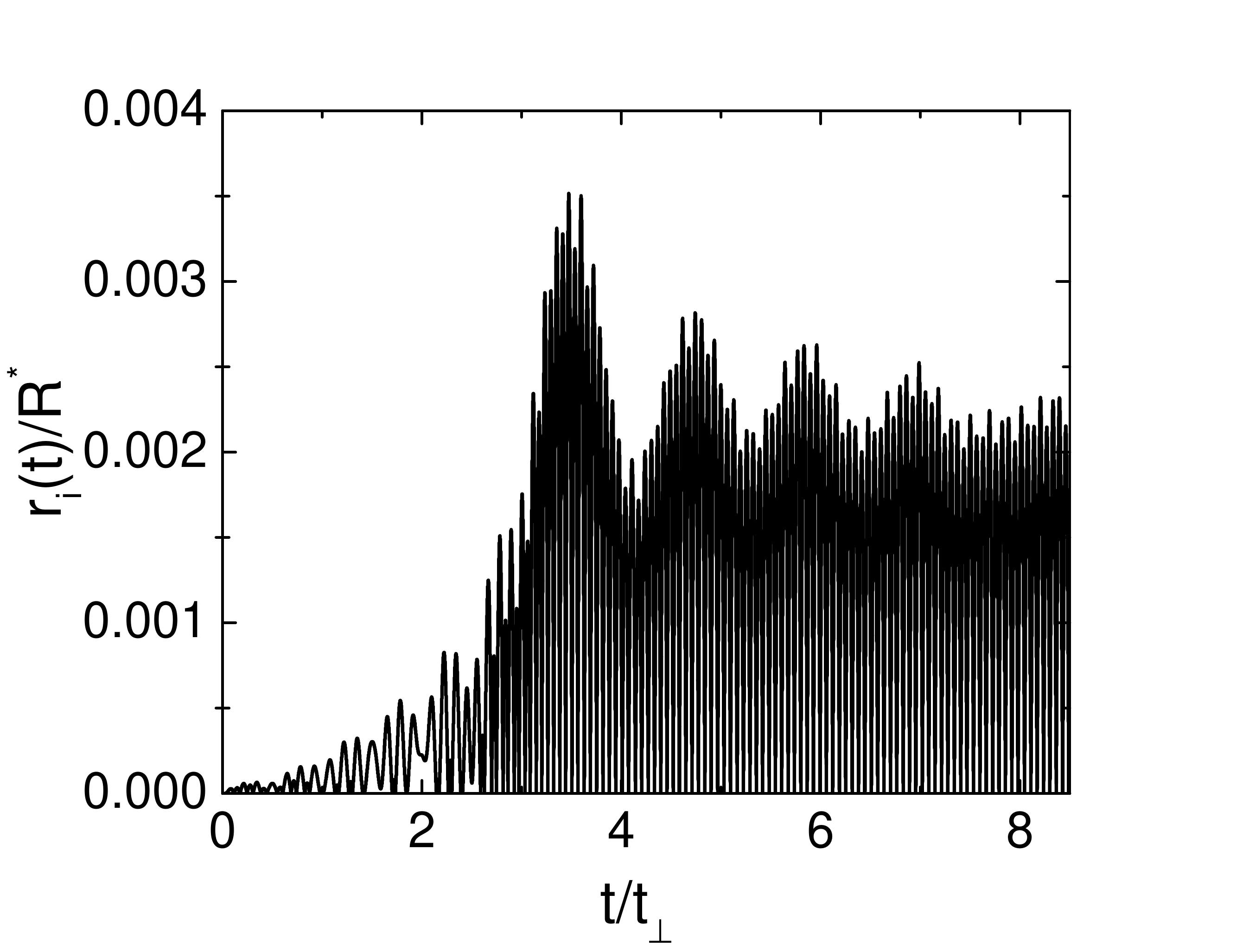}
\includegraphics[width=0.3\textwidth,height=0.3\textwidth]{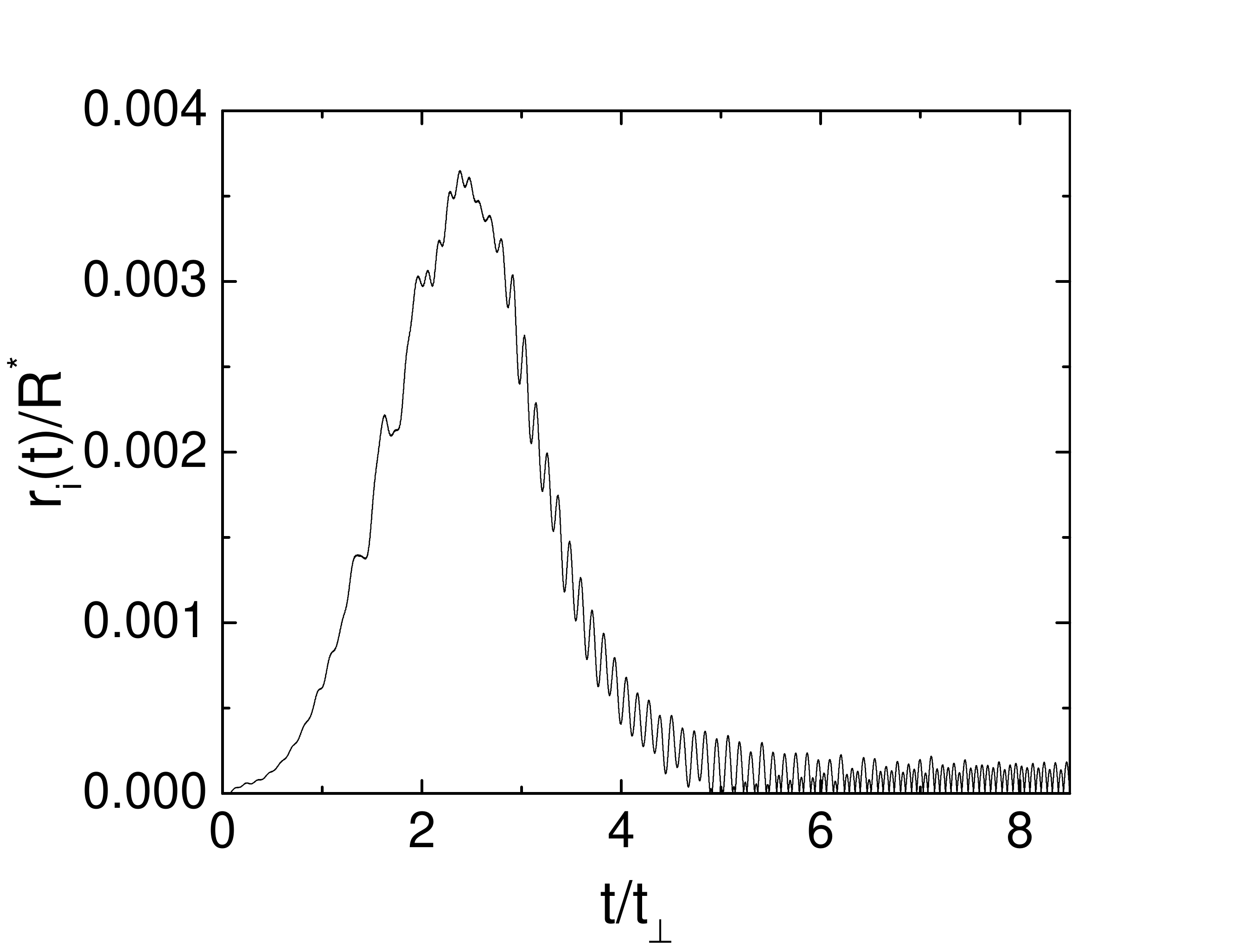}
\vspace{0.2cm}}
\caption{ (color online) The calculated atom scattering amplitude $f^+(t)$, transmission coefficient $T(t)$, coupling constant $g_{1D}(t)$ and the ion deviation from the center of the trap $r_i(t)$ for the ion being initially at rest (i.e. with zero energy before the collision with the atom) for three different values of the ratio $a_{\perp}/a_s$. This parameter fixes the coupling constant $g_{1D}$ (\ref{eq:g1D}) for three different cases: resonant atom-ion repulsion near the atom-ion CIR (left panel: $a_{\perp}/a_s=1.56$ at $b^2=0.179 , c^2=0.1$ in Eq.(\ref{eq:vai}), strong atom-ion repulsion (central panel: $a_{\perp}/a_s=2.64$ at $b^2=c^2=0.1$ ) and very weak atom-ion attraction (right panel: $a_{\perp}/a_s=2.64$ at $b^2=0.01 , c^2=0.06$).
}
\label{fig:Fig2}
\end{figure*}

 In Fig.\ref{fig:Fig2} the result of calculations of the scattering amplitude $f^+(a_{\perp}/a_s,t)$ (\ref{eq:overlap}), the transmission coefficient $T(a_{\perp}/a_s,t)$ (\ref{eq:T}), the effective coupling constant $g_{1D}(a_{\perp}/a_s,t)$ (\ref{eq:g1D}) and the deviation of the Yb$^+$ ion from the center of the Paul trap
\begin{align}
r_i(t)=\sqrt{x_i^2(t)+y_i^2(t)+z_i^2(t)}
\end{align}
are presented as a function of time for three different values $a_{\perp}/a_s$.

One can see  that after the collision which occurs at the time-interval $t_{\perp}\lesssim t \lesssim 4 t_{\perp}$ all the calculated scattering parameters and the value $r_i(t)$ reach stable regime at the times $t\sim 6t_{\perp}$ independently of the intensity of the atomic-ion interaction. The left graphs show the results of calculations at $a_{\perp}/a_s=1.56$ near the CIR, where the coupling constant diverges ($g_{1D}\simeq +65$) and total reflection is observed ($T\rightarrow 0$). We also see a very weak perturbation of the ion trajectory $ r_i (t) $ as a result of the collision and the absence of ion heating induced by micromotion. The reason for this is the "fermionization" of the relative atom-ion motion at the CIR point~\cite{Selim}: the atom-ion wave function is rearranged in such a way that its modulus squared at small atom-ion distances repeats the modulus squared  of the wave function of a pair of noninteracting fermions, i.e. the atom and the ion can not fully approach each other, what leads to $T(t\rightarrow\infty)\rightarrow 0$. This partially compensates the long-range character of the atom-ion interaction $V_{ai}(\mid \vek{r}_a-\vek{r}_i(t)\mid )$ and, as a consequence, prevents the ion micromotion-induced heating during collision. Actually, here the deviation of the Yb$^+$ ion from the center of the Paul trap after the collision does not exceed the value $~5\times10^{-5}R^*$.

The picture changes dramatically when we move from the resonance region around the CIR (region of "fermionization"). It is illustrated by the central and right graphs in Fig.\ref{fig:Fig2}. The central graphs illustrate the atom-ion collision with the repulsive atom-ion interaction $V_{ai}$ giving the ratio $a_{\perp}/a_s=2.64$ and the coupling constant $g_{1D}=1.82$. We observe that outside of "fermionization" the atom-ion interaction permits a closer approach between the atom and ion during collision ($T(t\rightarrow \infty) \simeq 0.1$) and significant micromotion-induced heating of the ion: the amplitude of deviation of the ion from the center of the Paul trap increases by more than an order of magnitude in comparison with the case of CIR, up to the value $\sim 2\times 10^{-3}R^*$. The right graphs illustrate the case of weak attraction between the atom and the ion $a_{\perp}/a_s=-260$ giving negative coupling constant $g_{1D}=-1.54$. Here, we also observe a considerable deviation of the ion from the center of the Paul trap after the collision, where $r_i(t\rightarrow\infty)$ approach the values $\sim 2\times 10^{-4}R^*$. This significant perturbation of the ion at close to zero atomic-ion scattering length $a_s\rightarrow -0$ (right panel of Fig.2) can be explained by the following way. The small value of the effective atom-ion attraction $a_s\rightarrow -0$ is partially compensated by its long-range nature $V_{ai}(\mid \vek{r}_i-\vek{r}_a\mid)\rightarrow -C_4/\mid \vek{r}_i-\vek{r}_a\mid^4$, which leads to significant perturbation $g_{1D}=-1.54$ of the ion at the moment of collision and to its noticeable heating as a result of the collision (see the time-dynamics of the quantity $r_i(t)$ in the right panel of Fig.2).

We have to note the visible oscillations in the coupling constant $g_{1D}(t)$  and in the ion deviation from the center of the Paul trap $r_i(t)$  at $a_{\perp}/a_s=1.56$ (i.e. near CIR). It is noteworthy that their time period $\sim 0.5 t_{\perp} = 0.5 (2\pi/\omega_{\perp})$ corresponds to the frequency $2\omega_{\perp}$ of the virtual transitions between the input channel $(n=0,I)$ to the closed first excited channel $(n=2,I)$, given by the transverse oscillations of the atom in the optical trap (see Fig\ref{fig:Fig2e}). This is consistent with the physical interpretation of the CIR as a Feshbach-like resonance in the first closed transverse channel~\cite{Bergeman}. Here, $n$ and $I$ define the quantum numbers of the atom in the atomic trap and the set of ion quantum numbers in the Paul trap correspondingly. When displaced from the resonance region (see the central and right panels in Fig.2), these oscillations of $g_{1D}(t)$ and $r_i(t)$ disappear, since the resonance conditions between the transverse quantum states of the atom are violated (see Fig\ref{fig:Fig2e}).
In this case, the higher-frequency oscillations of the values $g_{1D}(t)$ and $r_i(t)$ visible in the nonresonant cases are determined by the frequencies $\omega_i$ and $\Omega_{rf}$ of the Paul ion trap, which are much higher than the frequency of the atomic trap $\omega_{\perp}$.

The performed analysis demonstrates the suppression of micromotion-induced heating of an ion in a collision with a slow atom in the CIR region $a_{\perp}/a_s \simeq 1.5$ due to the effect of "fermionization" of their relative dynamics. In the next subsection, we explore the possibility of using this effect to enhance sympathetic cooling of ions by cold atoms in a hybrid atom-ion trap.

\subsection{Improving efficiency of sympathetic cooling in atom-ion confined collisions near CIR}

To analyze the possibility of improving sympathetic cooling of ions in hybrid atomic-ion traps (see Fig.1) with cold atoms, we have calculated the mean kinetic energy of the $^{171}$Yb$^+$ ion after collision with a cold $^6$Li atom in such a trap
\begin{align}
\label{eq:energii}
\langle E_i^{(out)}\rangle = \frac{1}{t_{max}-t_{out}}\int_{t_{out}}^{t_{max}}E_i(t)dt \,,
\end{align}
depending on the parameter $a_{\perp}/a_s$ near CIR and outside the resonant area. In the above formula, the ion energy $E_i(t)=\vek{p}_i^2(t)/(2m_i)$ was calculated by integrating a coupled system of equations (\ref{eq:Schr}),(\ref{eq:Hamilton}). The limit of integration of the system of equations $t_{max}$ was chosen from the condition of the calculated parameters reaching stable values after the collision (see the previous subsection), in the region $t\gtrsim 6t_{\perp}$. The lower limit $t_{out}$ of integration in formula (\ref{eq:energii}) was chosen in a similar way. In the calculation the following values $t_{out}=9t_{\perp}$ and $t_{max}=10t_{\perp}$ for these parameters were used.

The initial conditions for a cold atom were chosen similarly to the previous subsection. The initial mean ion energy
\begin{align}
\label{eq:energiin}
\langle E_i^{(in)}\rangle = \frac{1}{t_{in}}\int_{0}^{t_{in}}E_i(t)dt =0.33E^*\,,
\end{align}
was chosen significantly exceeding the longitudinal $E_{coll}=0.004E^*$ as well as transverse $E_{\perp}=\hbar\omega_{\perp}=0.02E^*$ energy of the atom. The upper limit at calculating the initial mean ion energy $t_{in}\lesssim t_{\perp}$ was chosen from the region before the collision, where an atom and an ion were not interacting $V_{ai}(\mid \vek{r}_a-\vek{r}_i(t_{in})\mid)\rightarrow 0$.
Here we have considered two fundamentally different cases which however correspond to the same mean initial energy of the ion $\langle E_i^{(in)} \rangle$.

The first case: in the initial state, the ion has only one transverse momentum component
\begin{align}
\label{eq:2d}
\vek{r}_i(t=0) &= 0 \nonumber\\
p_{i,x}(t=0) &= \sqrt{2 m_i E_{ix}^{(0)}},\nonumber\\
p_{i,y}(t=0) &= 0,\nonumber\\
p_{i,z}(t=0) &= \sqrt{2 m_i E_{iz}^{(0)}}\,,
\end{align}
which leads to a “head-on collision” of an ion oscillating in one xz plane with an incident atom moving along the z axis. In our calculations $E_{ix}^{(0)}=E_{iz}^{(0)}$ were chosen equal to $0.0625E^*$, which gives the initial mean energy $\langle E_i^{(in)}\rangle=0.33E^* \gg E_{\perp}+E_{coll} =0.024E^*$.

The second case: not a head-on collision with the same mean initial ion energy, which was simulated by the following initial conditions for the ion
\begin{align}
\label{eq:3d}
x_{i}(t=0) &= x_{i0}\nonumber\\
y_{i}(t=0) &= y_{i0},\nonumber\\
z_{i}(t=0) &= z_{i0},\nonumber\\
\vek{p}_i(t=0) &= 0 \,,
\end{align}
specifying its initial position as $x_{i0}=z_{i0}=0.3R^*$, $y_{i0}=0.6R^*$. In this, case the initial condition generates 3D motion of the ion in the Paul trap before the collision.
\begin{figure}
\centering\includegraphics[scale=0.3]{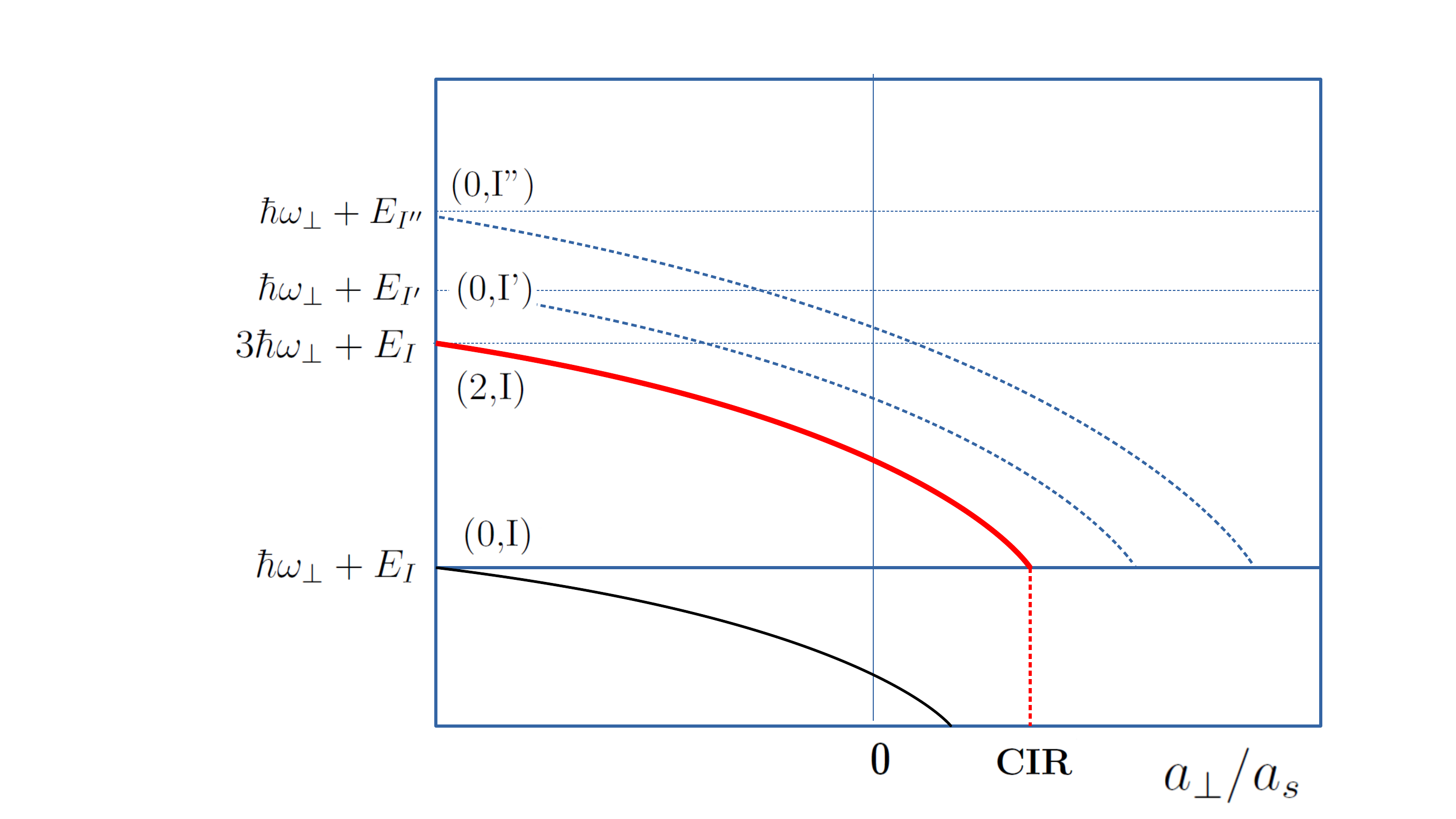}
\caption{(color online) Schematic representation of the spectrum of the atom-ion system confined in hybrid atom-ion trap as a function of the ratio $a_{\perp}/a_s$. The pair $(n,I)$ indicates the atom quantum number $n$ and the set of ion quantum numbers $I$. The point of the cross of the energy curve of the first excited state with respect to the atomic motion $(2,I)$ with the threshold of the entrance channel of the system $(0,I)$ defines the position of the CIR on the $a_{\perp}/a_s$-axis.}
\label{fig:Fig2e}
\end{figure}
\begin{figure}
\parbox{0.5\textwidth}{
\centering\includegraphics[scale=0.3]{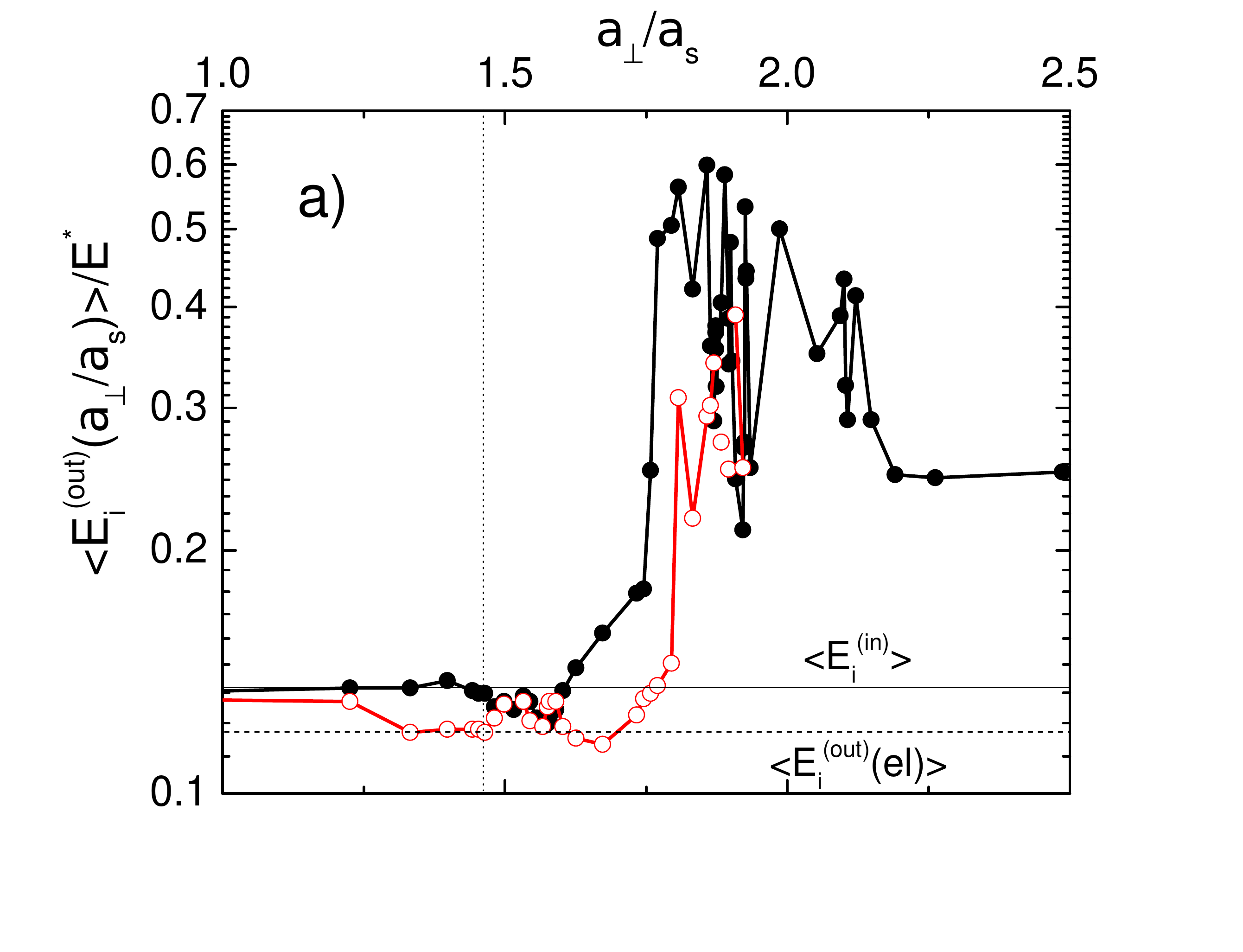}
\vspace{-1.2cm}}
\parbox{0.5\textwidth}{
\centering\includegraphics[scale=0.3]{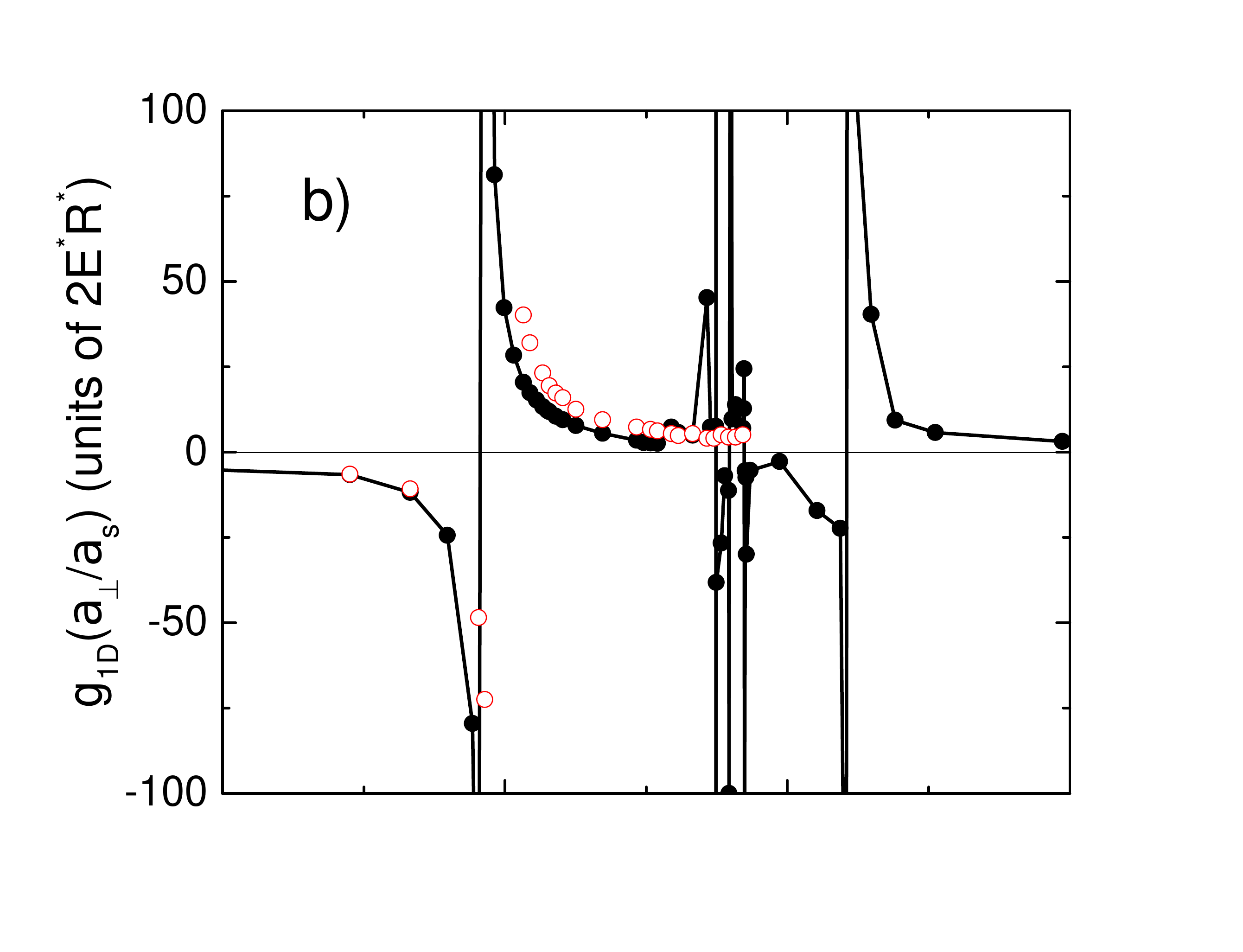}
\vspace{-1.4cm}}
\parbox{0.5\textwidth}{
\centering\includegraphics[scale=0.3]{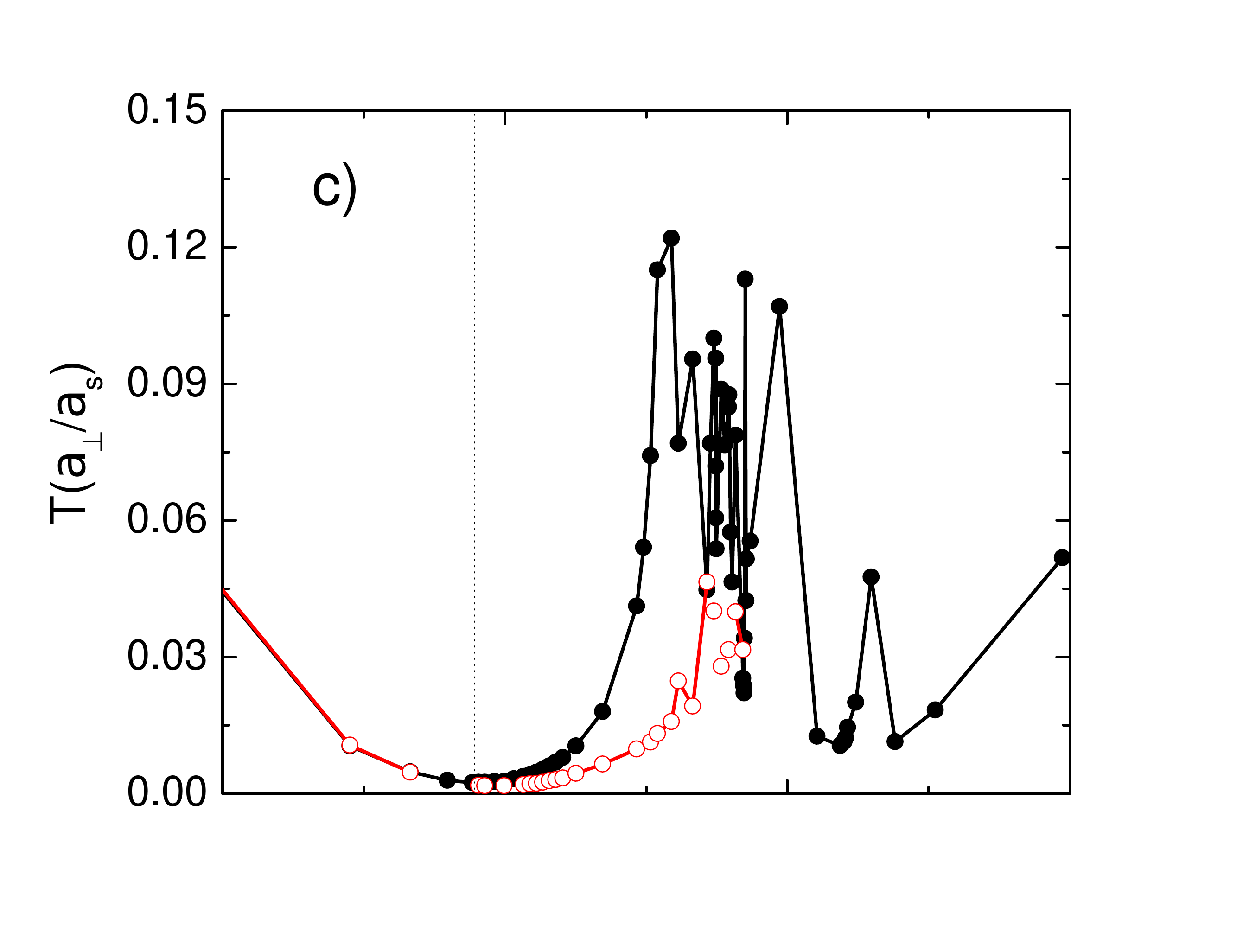}
\vspace{-1.4cm}}
\parbox{0.5\textwidth}{
\centering\includegraphics[scale=0.3]{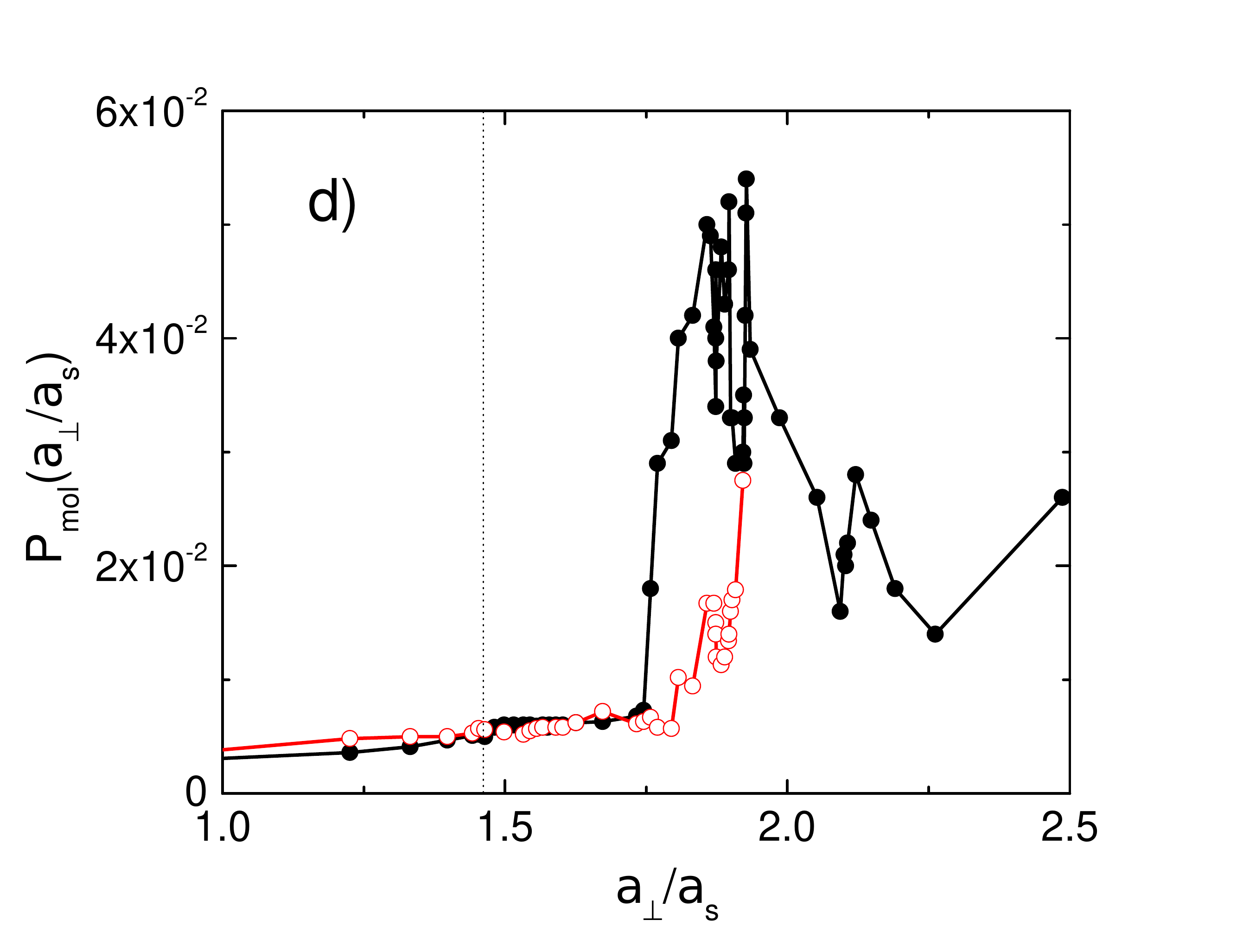}}
\caption{(color online) The calculated mean ion energy $\langle |E^{(out)}_{ik}|\rangle$ after atom-ion collision, effective coupling constant $g_{1D}$, transmission coefficient $T$ and the molecule formation probability $P_{mol}$ as a function of the ratio $a_{\perp}/a_s$. The black circles connected by the solid black lines indicate the results of calculations performed for the "head-on collisions" with the initial conditions for the ion (\ref{eq:2d}). The open circles are connected by the solid red lines related to "not head-on collisions" obtained with the initial conditions for the ion(\ref{eq:3d}).}
\label{fig:Fig3}
\end{figure}

The results of calculations of the final mean ion energy $\langle E_i^{(out)}\rangle$ (\ref{eq:energii}) after the collision with the cold atom for these two cases are given in Fig.\ref{fig:Fig3} as a function of the ratio $a_{\perp}/a_s$. In this figure the calculated coupling constant $g_{1D}$ (\ref{eq:g1D}), the transmission coefficient $T$ (\ref{eq:T}) and the molecule formation probability $P_{mol}$ (defined below by Eq.(\ref{eq:mol})) corresponding to the same values $a_{\perp}/a_s$ are also presented. One can see that the calculated curves $\langle E_i^{(out)}(a_{\perp}/a_s)\rangle$ presented in Fig.\ref{fig:Fig3}a have the region of minimal values around the point of the CIR where $a_{\perp}/a_s=1.55$, $g_{1D} \rightarrow \pm\infty$ (Fig.\ref{fig:Fig3}b) and the transmission $T\rightarrow 0$ (Fig.\ref{fig:Fig3}c) in agreement with the result of the previous subsection obtained for the case when the ion was initially at rest in the center of the Paul trap. In Fig.\ref{fig:Fig3}a, the initial mean energy of the ion $\langle E_i^{(in)}\rangle$ together with the value for the final ion energy $\langle E_i^{(out)}(el)\rangle =\langle E_i^{(in)}\rangle-\Delta E_i(el)$, calculated according to the classical formula
\begin{align}
\label{eq:ellastic}
\Delta E_i(el)= \frac{4m_a m_i}{(m_a+m_i)^2}(\langle E_i^{(in)}\rangle-E_{coll}-E_{\perp})\,
\end{align}
and valid for the central elastic collision of two classical particles, are also presented. The performed investigation demonstrates
the existence of rather broad region $1.3 \lesssim a_{\perp}/a_s \lesssim 1.7$ where the ion lost energy during not a head-on collision with the atom approaches the value following from the classical consideration (\ref{eq:ellastic}). This region, however, narrows significantly in the case of  “head-on collision”.

With the growth of the ratio $a_{\perp}/a_s$ we get into the area of the resonance heating of an ion in a collision with a cold atom (see Fig.\ref{fig:Fig3}). The calculated curves of $g_{1D}(a_{\perp}/a_s)$, $T(a_{\perp}/a_s$ and $P_{mol}(a_{\perp}/a_s)$ presented in the graphs Figs.\ref{fig:Fig3}b-d allow us to interpret this effect. In Fig.\ref{fig:Fig3}b we see a series of very narrow resonances in the coupling constant $g_{1D}$, where the edge left of them is the classical broad CIR due to the virtual transition $(n=0,I)\rightarrow (n=2, I)$ from the atomic ground transverse ($n=0$) to the closed transverse first excited state ($n=2$) without change of the ion state ($I$) in the Paul trap (see Fig\ref{fig:Fig2e}). However, it is known~\cite{Thorwart,Melezhik2009,Sala} that in the case of confined collision of distinguishable particles the CIR is splitting due the coupling upon collision of the relative and center-of-mass motions, which leads to the transitions $(0,I)\rightarrow (0,I'),(0,I'')...$ when approaching the resonant conditions with excitation of the center-of mass motion of forming molecular ion and ion excitation.
An indication of such transitions can be a resonant increase in the rate of formation of molecular ions with the transfer of the released energy to the excitation of the center-of-mass of the molecular ion~\cite{Melezhik2009,Sala}. To evaluate this effect, we have calculated the probability of the molecular ion formation by the formula
\begin{align}
\label{eq:mol}
P_{mol}= \int_0^{R^*}\psi^2(\vek{r}_a, t\rightarrow \infty)d\vek{r}_a\,,
\end{align}
which represents the probability for the atom to stick to the ion near the center of the Paul trap in the region $\lesssim R^*$ after collision, i.e. the probability of the formation of the molecular ion LiYb$^+$. The calculated curve $P_{mol}(a_{\perp}/a_s)$ given in Fig.\ref{fig:Fig3}d repeats the resonance behavior of the curves $g_{1D}(a_{\perp}/a_s)$ and $T(a_{\perp}/a_s)$ with the growth of the ratio $a_{\perp}/a_s$ right of the CIR position. We suppose that it is clear confirmation of the resonant mechanism of the molecular formation with transferring the energy release to the molecular center-of-mass excitation and ion heating developed in this region in Fig.\ref{fig:Fig3}a~\cite{Melezhik2009,Sala}.

\begin{figure*}
\parbox{0.95\textwidth}{
\includegraphics[width=0.3\textwidth,height=0.3\textwidth]{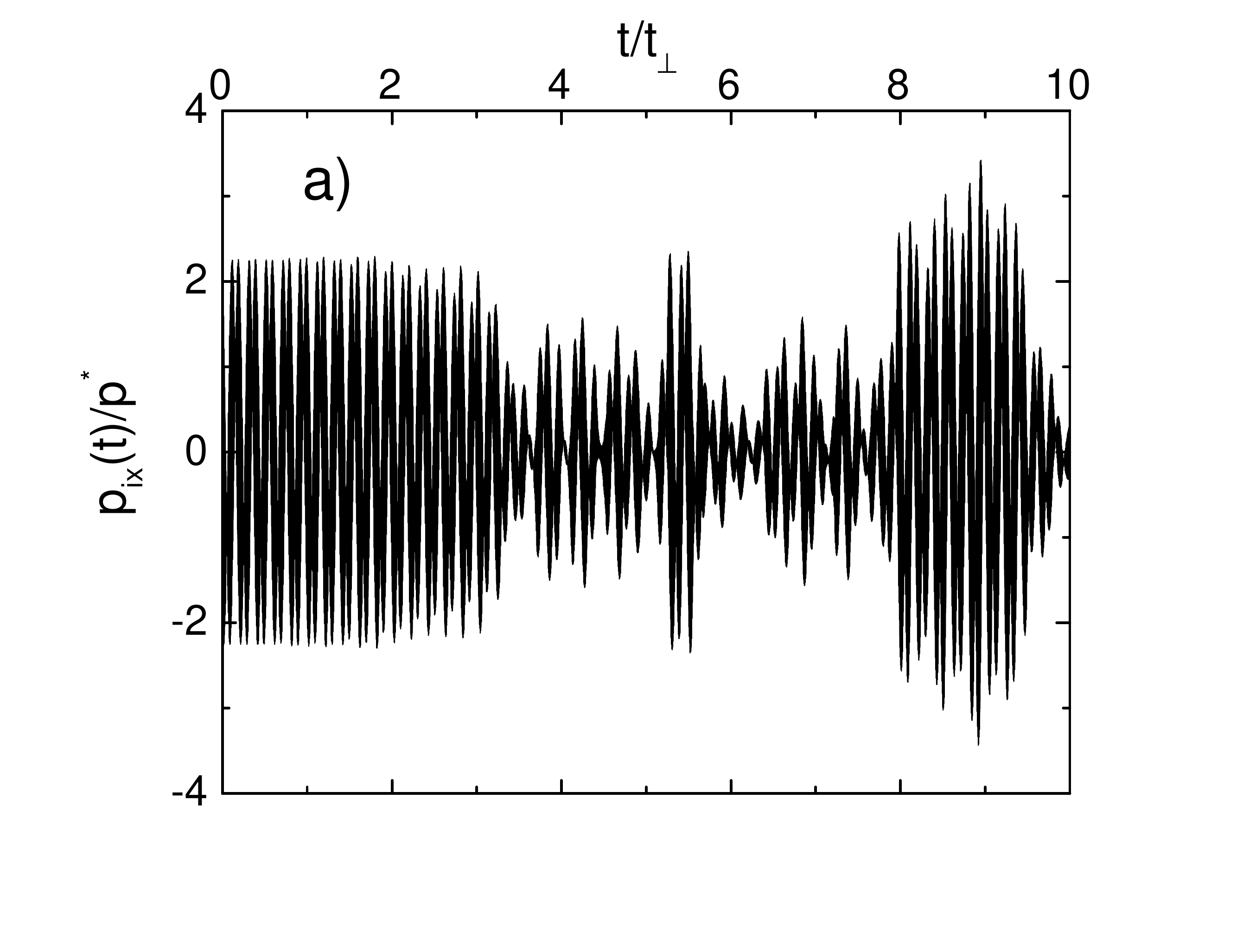}
\includegraphics[width=0.3\textwidth,height=0.3\textwidth]{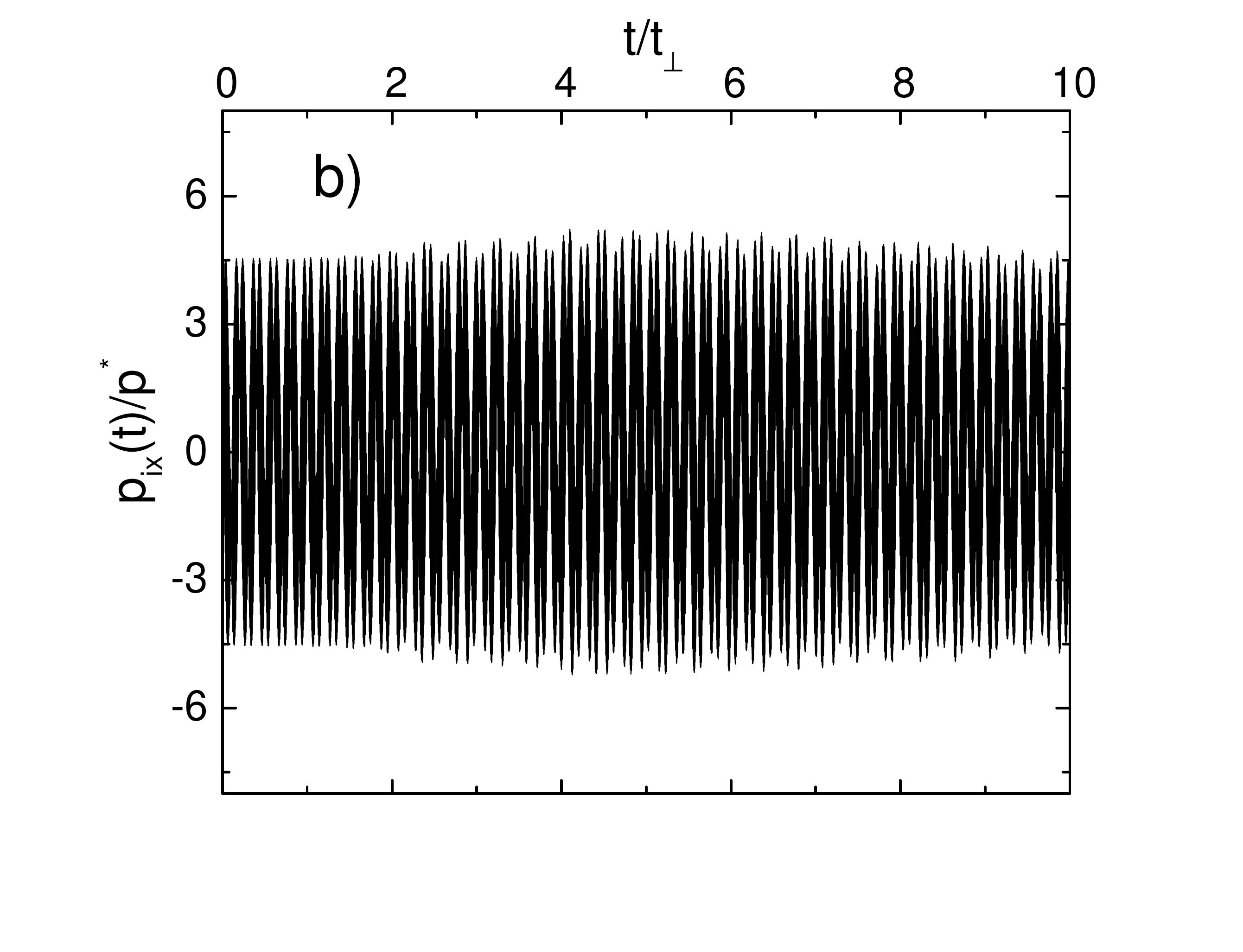}
\includegraphics[width=0.3\textwidth,height=0.3\textwidth]{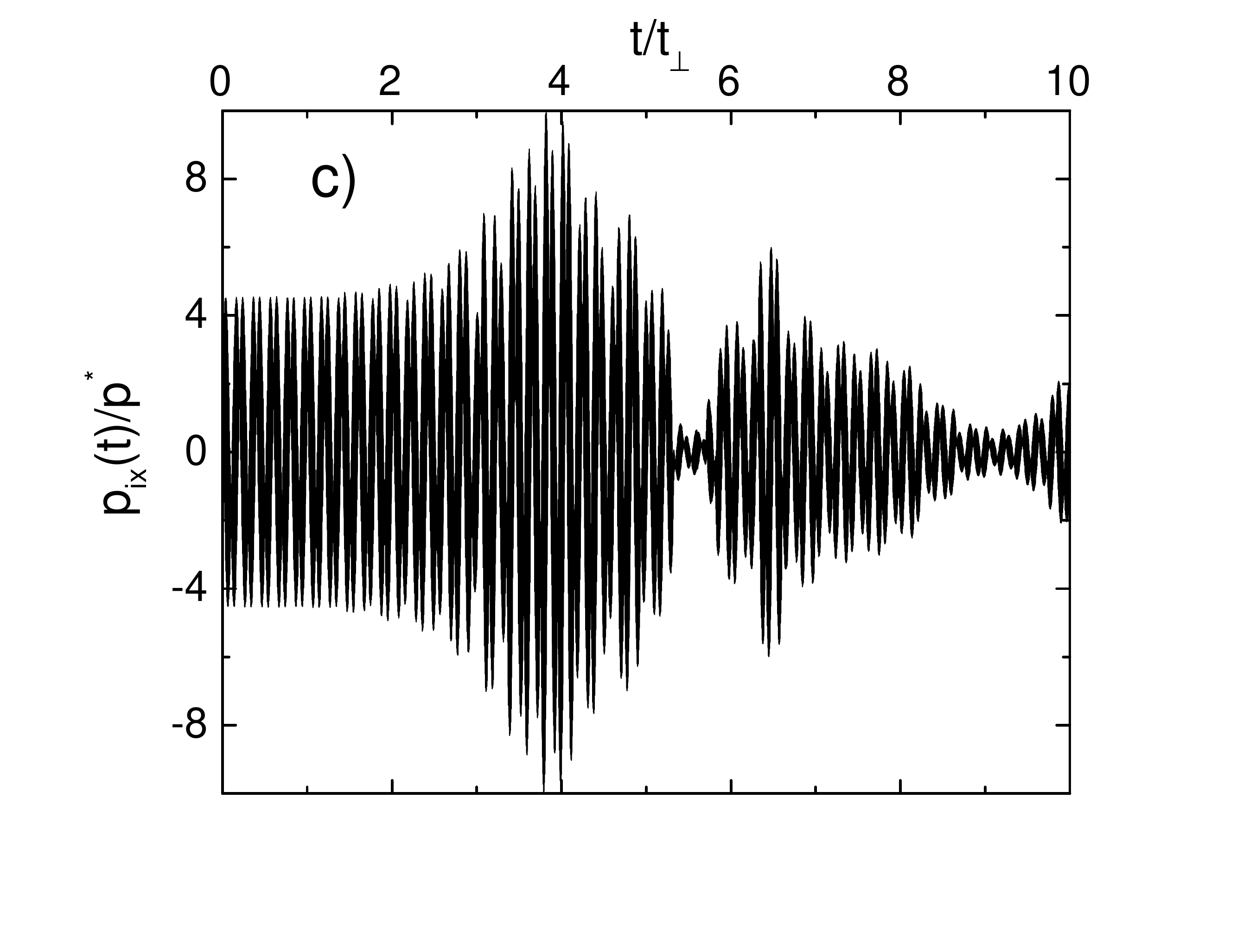}
\vspace{-1.2cm}}
\parbox{0.95\textwidth}{
\includegraphics[width=0.3\textwidth,height=0.3\textwidth]{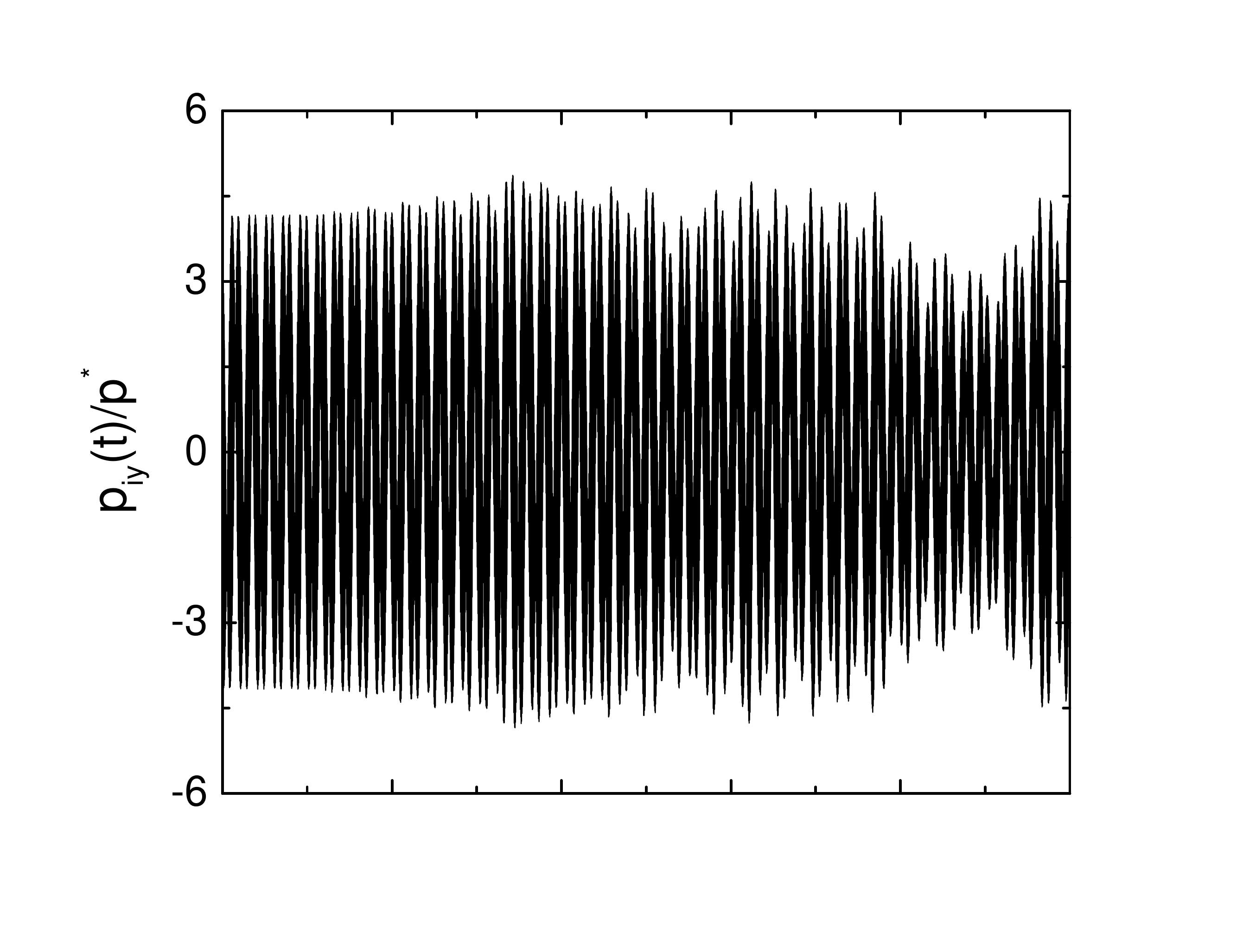}
\includegraphics[width=0.3\textwidth,height=0.3\textwidth]{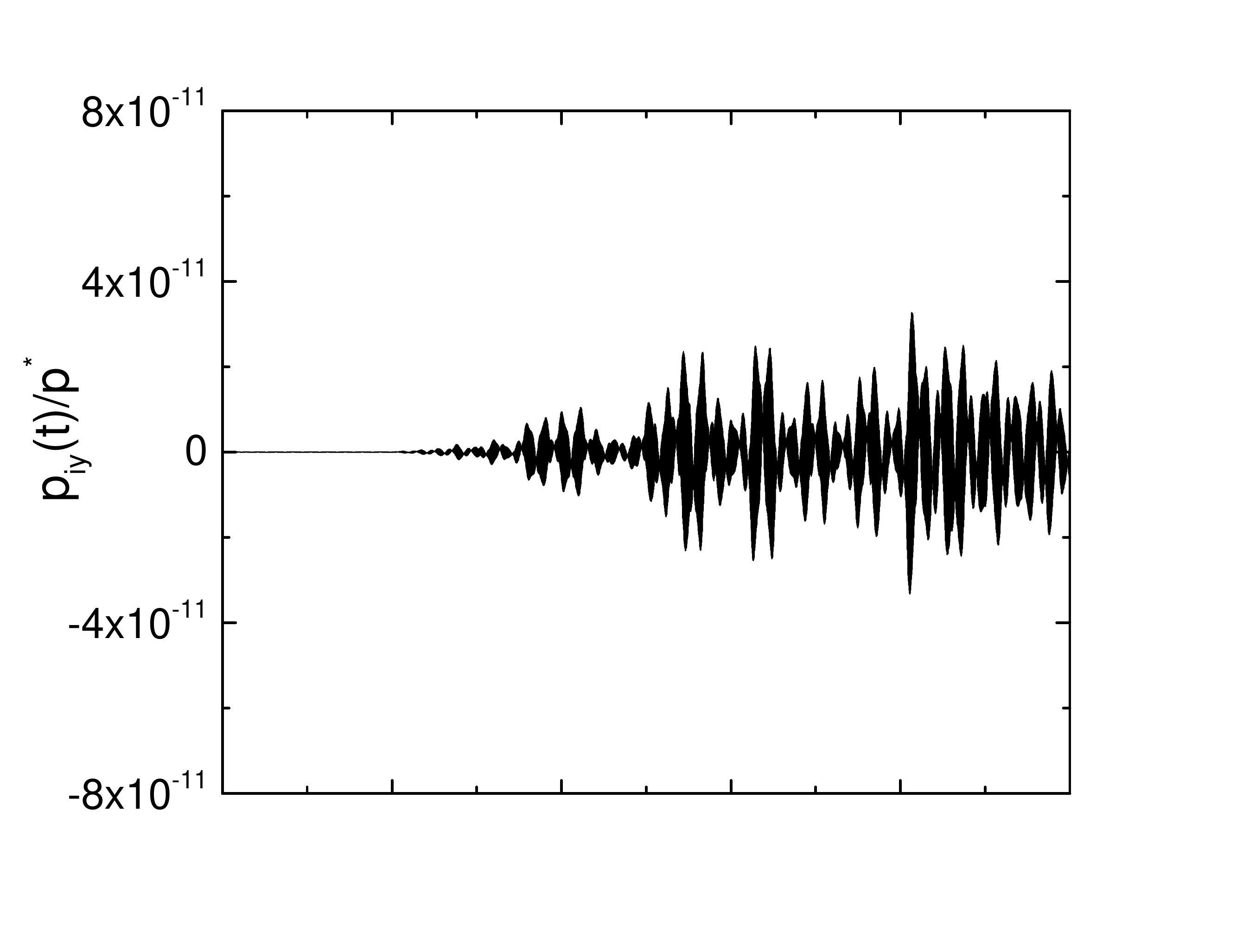}
\includegraphics[width=0.3\textwidth,height=0.3\textwidth]{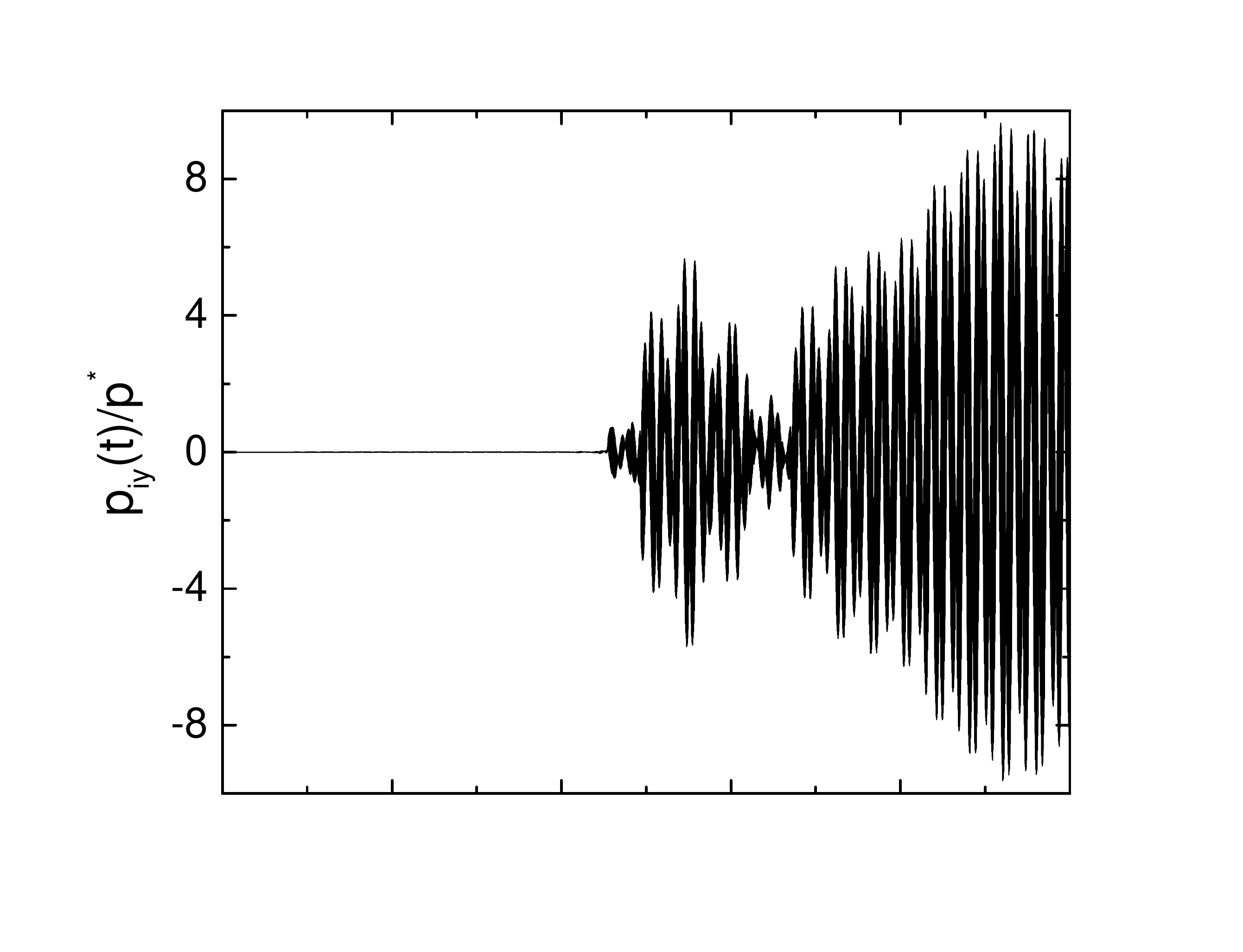}
\vspace{-1.2cm}}
\parbox{0.95\textwidth}{
\includegraphics[width=0.3\textwidth,height=0.3\textwidth]{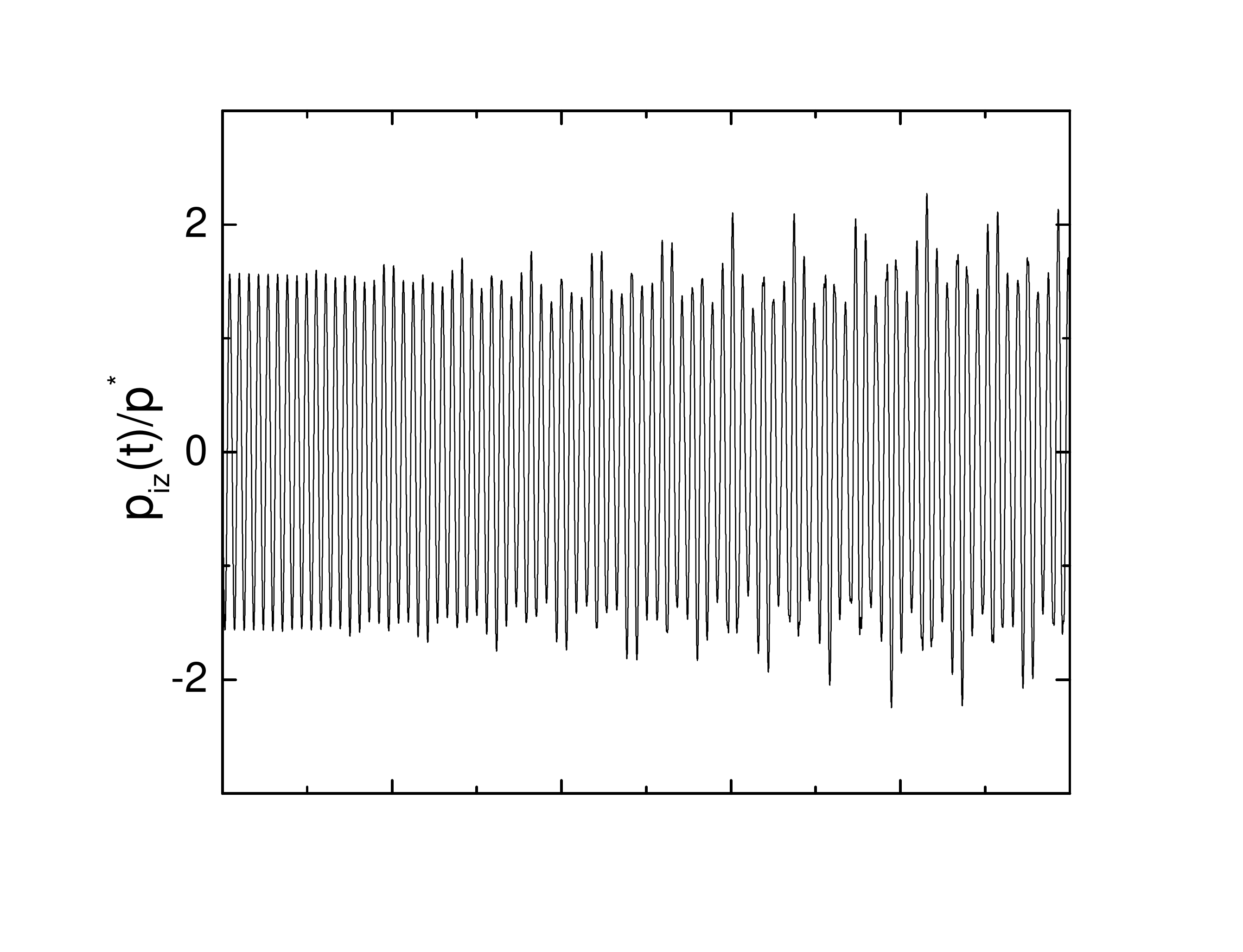}
\includegraphics[width=0.3\textwidth,height=0.3\textwidth]{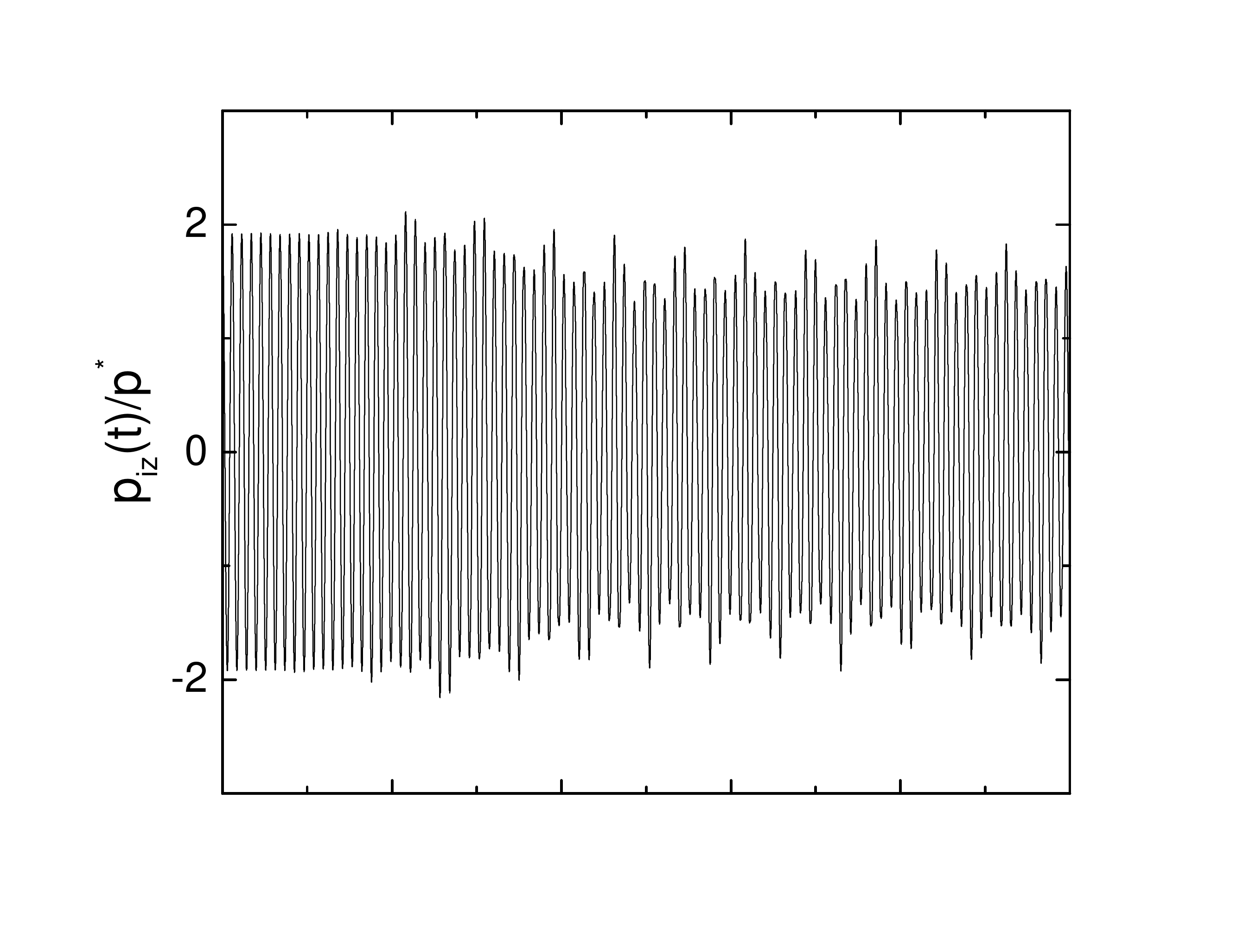}
\includegraphics[width=0.3\textwidth,height=0.3\textwidth]{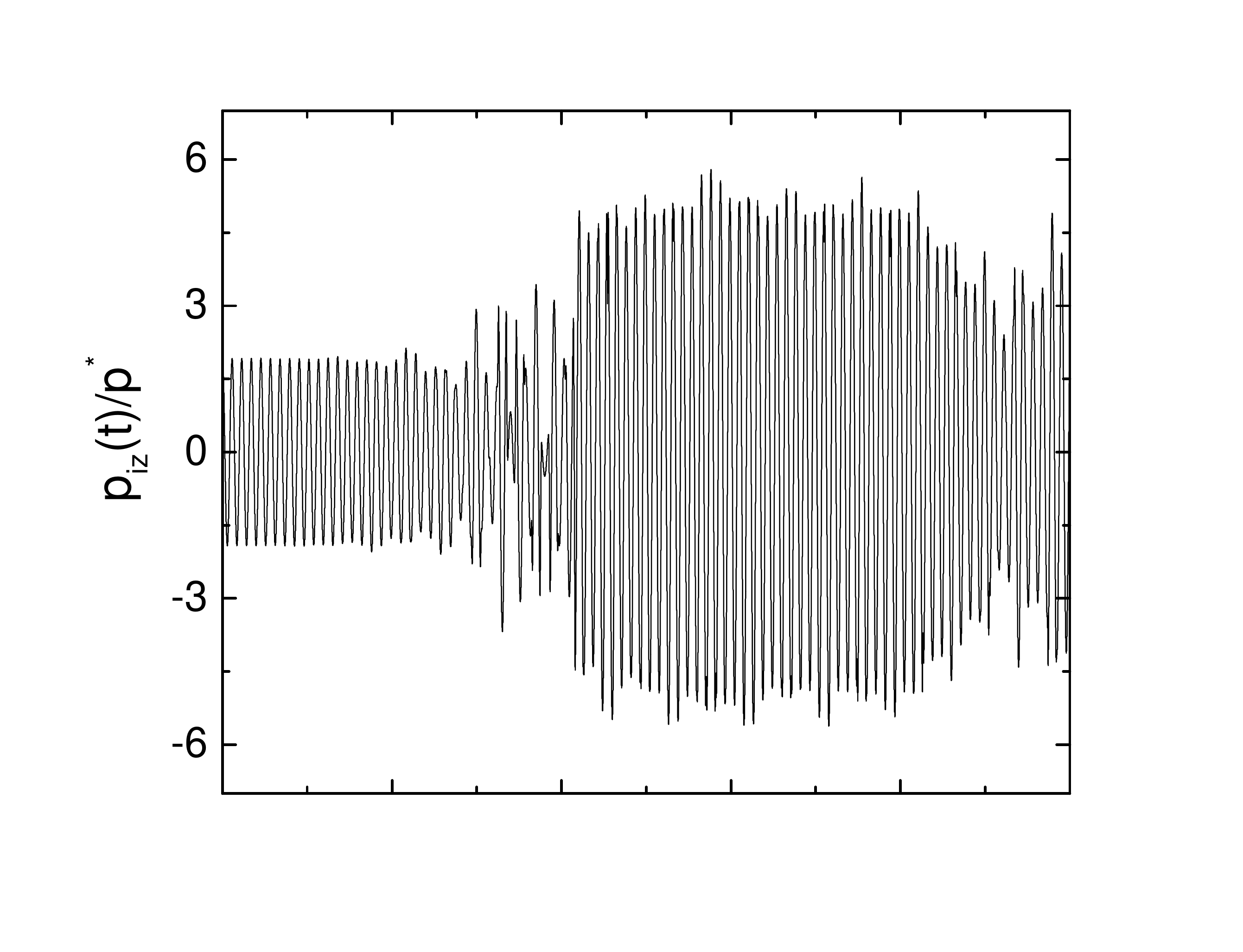}
\vspace{-1.2cm}}
\parbox{0.95\textwidth}{
\includegraphics[width=0.3\textwidth,height=0.3\textwidth]{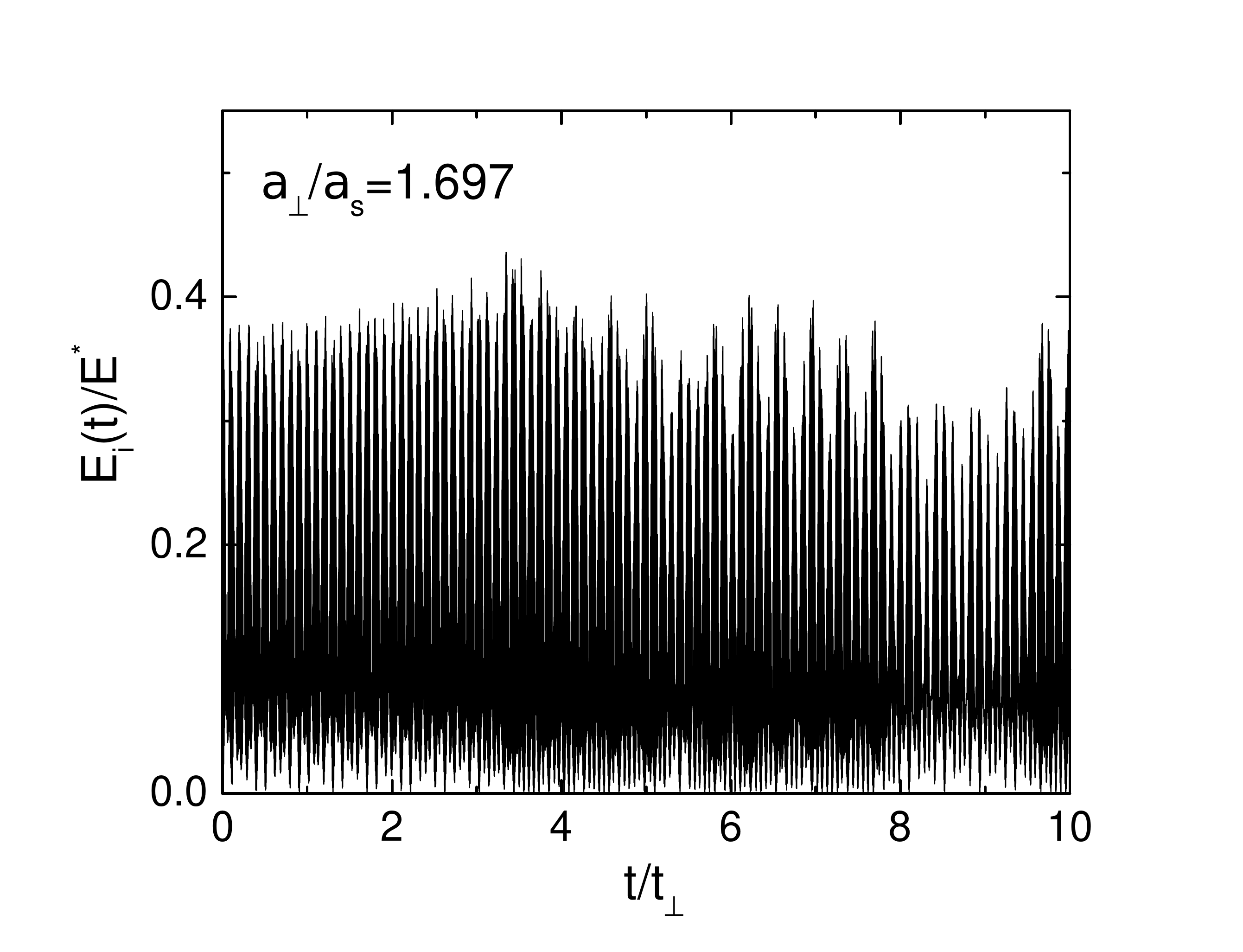}
\includegraphics[width=0.3\textwidth,height=0.3\textwidth]{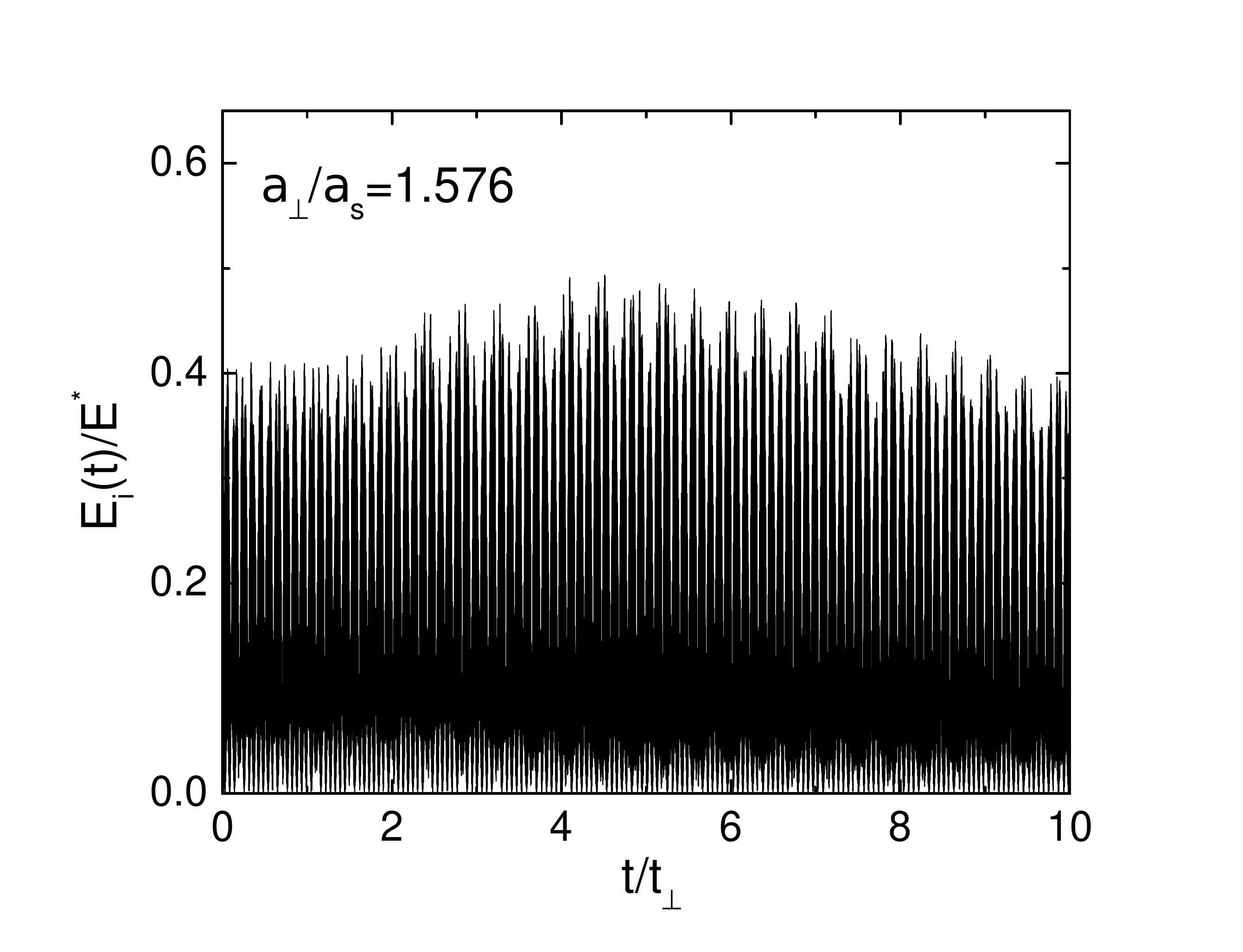}
\includegraphics[width=0.3\textwidth,height=0.3\textwidth]{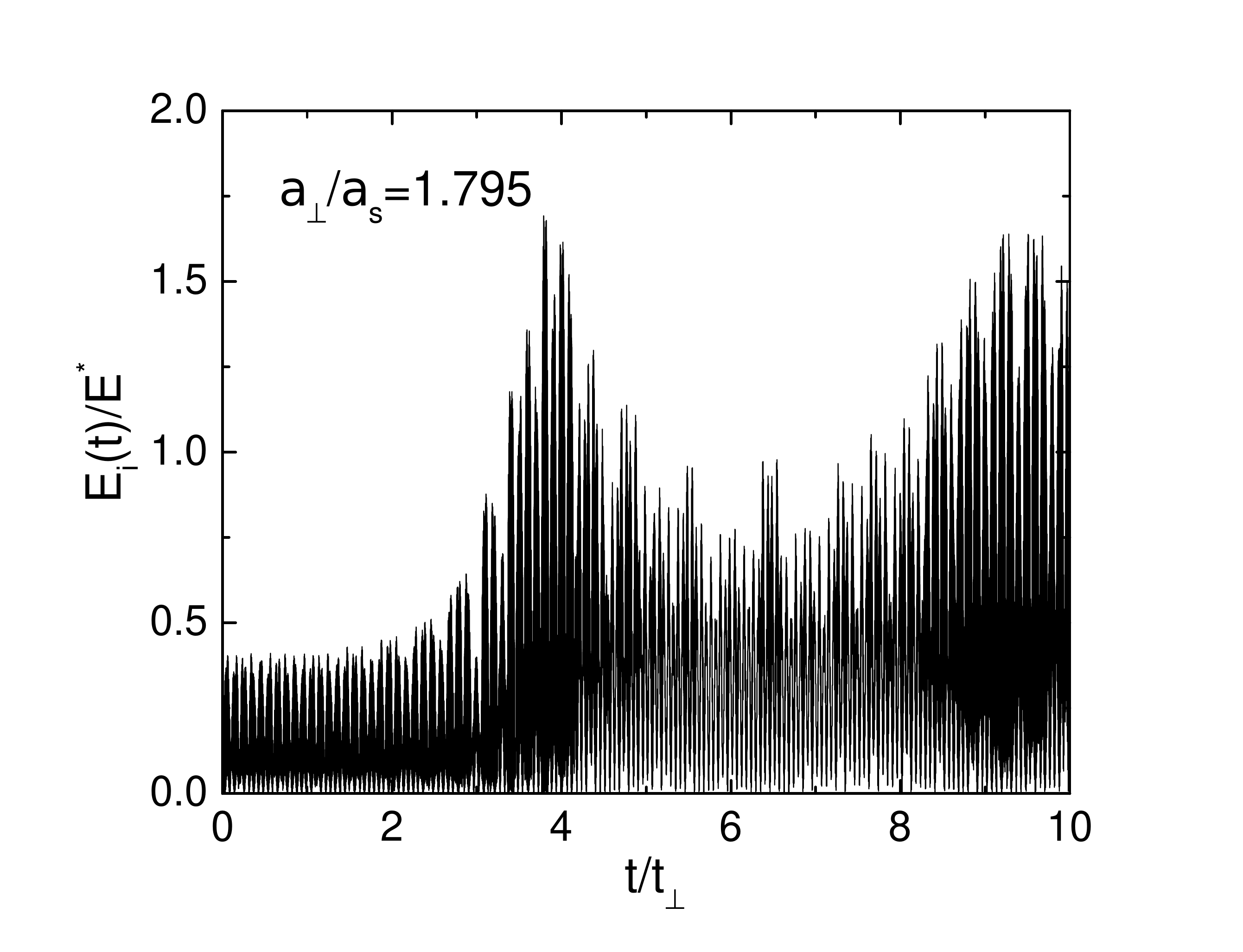}
\vspace{0.2cm}}
\caption{The calculated time-evolution of the ion momentum $\vek{p}_i(t)$ and its kinetic energy $E_i(t)$ during the collision with the atom. The left column of graphs (graphs a)) presents the result of the calculation with the potential of atomic-ion interaction $V_{ai}$ fixing the ratio $a_{\perp}/a_s=1.697$ slightly to the right of the CIR and with the initial condition (\ref{eq:3d}) specifying the 3D-motion of the ion in the Paul trap before the collision. The results presented in the central and right columns of the graphs (graphs b) and c)) were obtained with the initial condition (\ref{eq:2d}) specifying the 2D-motion of the ion in the Paul trap before the collision. The results shown in the central graphs (graphs b)) were obtained with the potential $V_{ai}$ giving the ratio $a_{\perp}/a_s=1.576$ in the resonance region slightly to the right of the CIR. The right graphs (graphs c)) illustrate the dynamics of the ion outside the resonance region for $a_{\perp}/a_s=1.796$.
}
\label{fig:Fig3a}
\end{figure*}

Figure \ref{fig:Fig3a} illustrates the time dynamics of an ion under sympathetic cooling and heating. It is shown that the ion performing 3D oscillations in the Paul trap before the collision at $a_{\perp}/a_s=1.697$ noticeably decreases its mean energy as a result of the collision with a cold atom slightly to the right of the CIR (bottom left graph in Fig.\ref{fig:Fig3a}). As we can see from the left column of the graphs, this occurs due to a decrease in its amplitude of oscillations in the transverse directions after the collision. The amplitude of oscillations in the longitudinal direction remains almost unchanged. There is also a decrease in the mean energy of the ion,  oscillating in the plane before the collision, after collision with a cold atom at $a_{\perp}=1.576$ slightly to the right of CIR (see central column of the graphs in Fig.\ref{fig:Fig3a}). In this case, its oscillations remain two-dimensional, the orientation of the plane of oscillations does not change, and the energy decreases due to a decrease in the amplitude of longitudinal vibrations. The graphs from the right column in Fig.\ref{fig:Fig3a} demonstrate the heating of an ion upon collision with a cold atom at $a_{\perp}/a_s=1.796$ outside the CIR region. As a result of the collision, the 2D oscillations of the ion before the collision  become 3D after the collision.

Thus, the performed analysis demonstrates that one can control the sympathetic cooling of ions in a hybrid atom-ion trap by tuning the ratio $a_{\perp}/a_s$ to the resonant region around the CIR, where cooling can reach the most optimal conditions. In the cases considered, the maximum loss of energy by an ion reaches the energy corresponding to the loss of energy by a heavy ball with a mass of $^{171}$Yb$^+$ in its elastic head-on collision with a light slow ball with a mass of $^6$Li.

\subsection{Sympathetic cooling in atom-ion confined collisions in secular approximation}

Here, we investigate the effect of a CIR on the efficiency of sympathetic cooling in a hybrid atomic-ion trap in the framework of the time-independent secular approximation~\cite{LeibfriedRMP03,Melezhik2019} with constant frequencies for the ion-trap interaction (\ref{eq:Uion}). Since the secular approximation for the potential of interaction of an ion with a trap does not contain the RF part and, therefore, does not cause micromotion of an ion, the calculation within its framework allows one to estimate the effect of micromotion-induced  heating at the collision with a cold atom by comparing  with the result of the previous subsection including the micromotion.
\begin{figure}
\parbox{0.5\textwidth}{
\centering\includegraphics[scale=0.3]{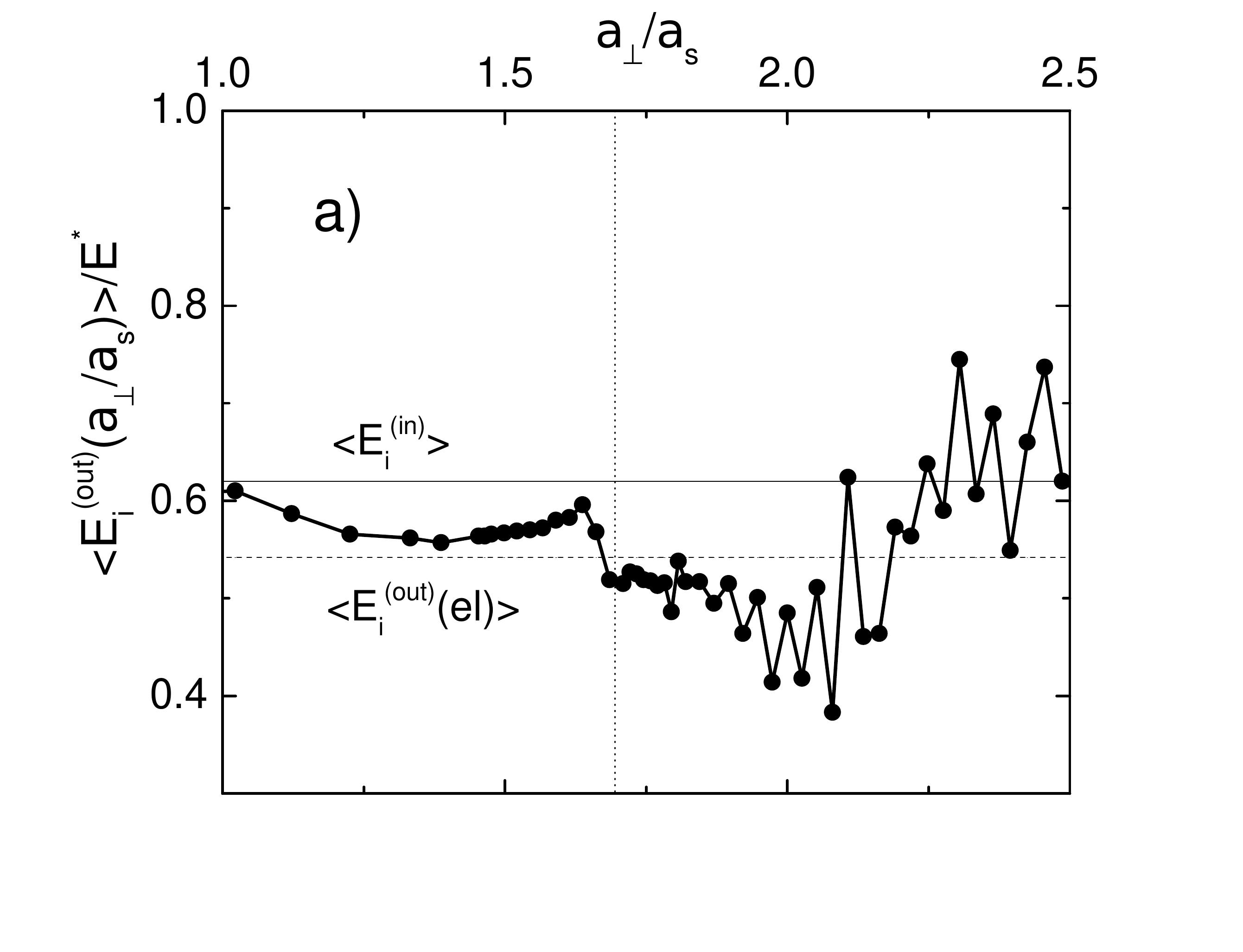}
\vspace{-1.2cm}}
\parbox{0.5\textwidth}{
\centering\includegraphics[scale=0.3]{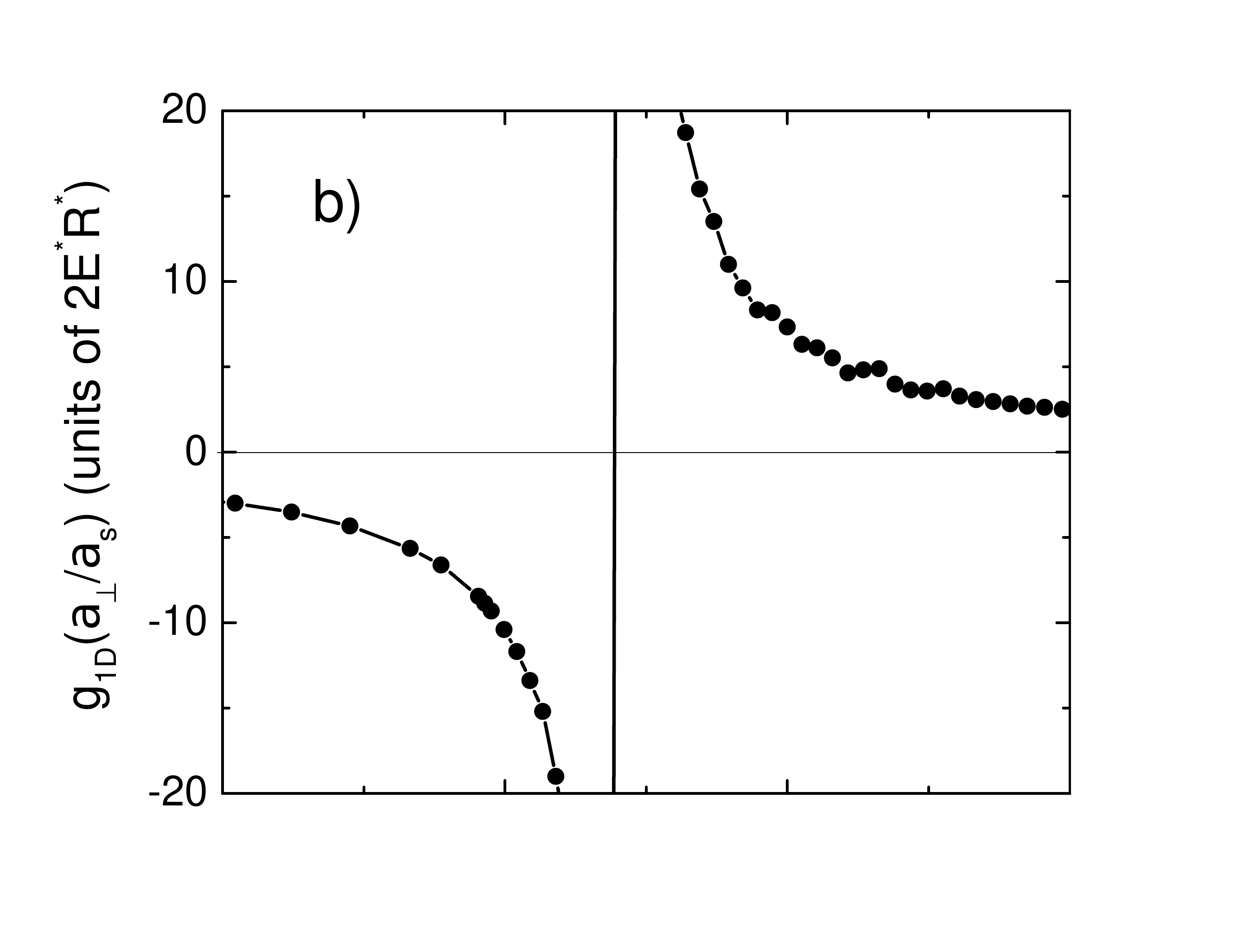}
\vspace{-1.4cm}}
\parbox{0.5\textwidth}{
\centering\includegraphics[scale=0.3]{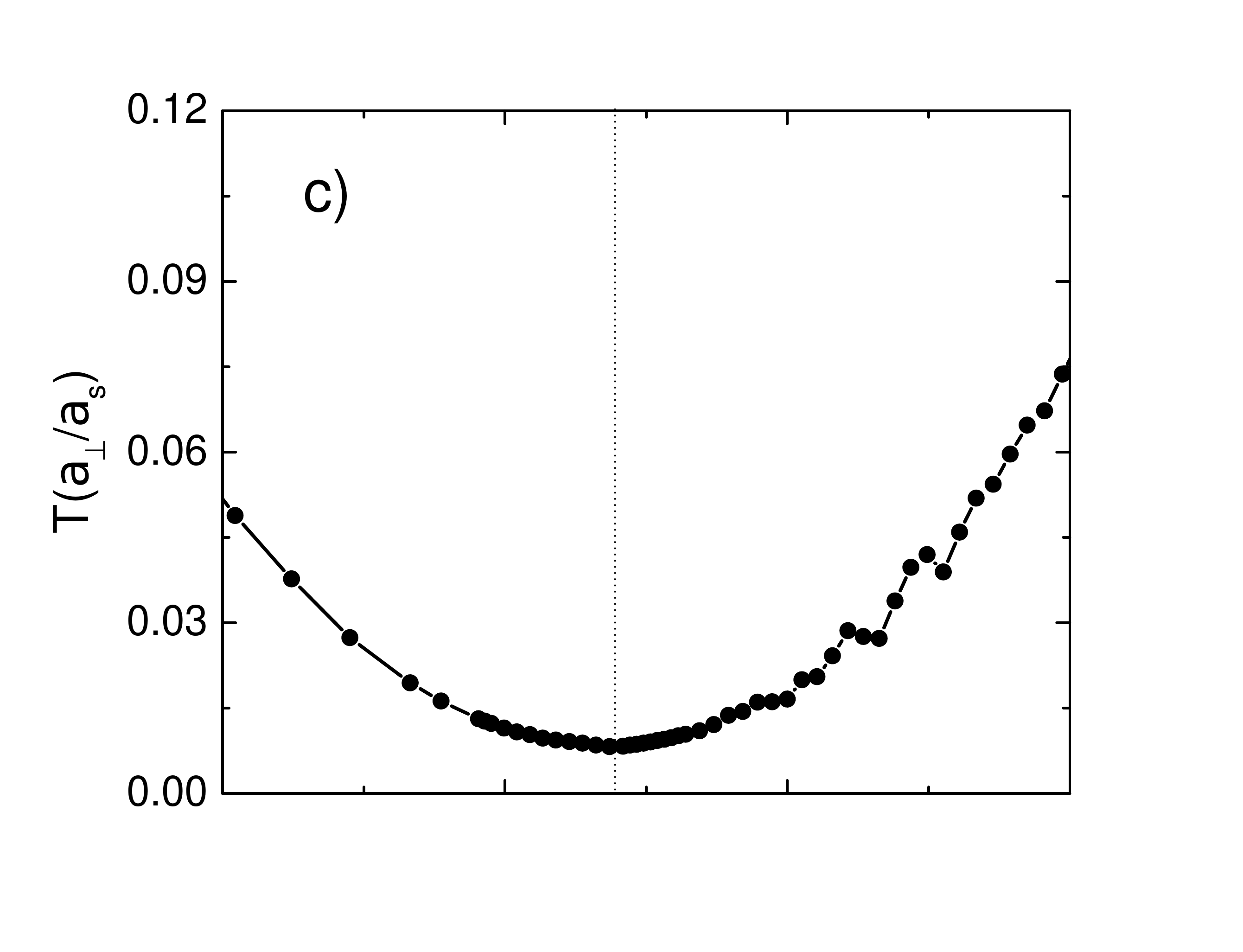}
\vspace{-1.4cm}}
\parbox{0.5\textwidth}{
\centering\includegraphics[scale=0.3]{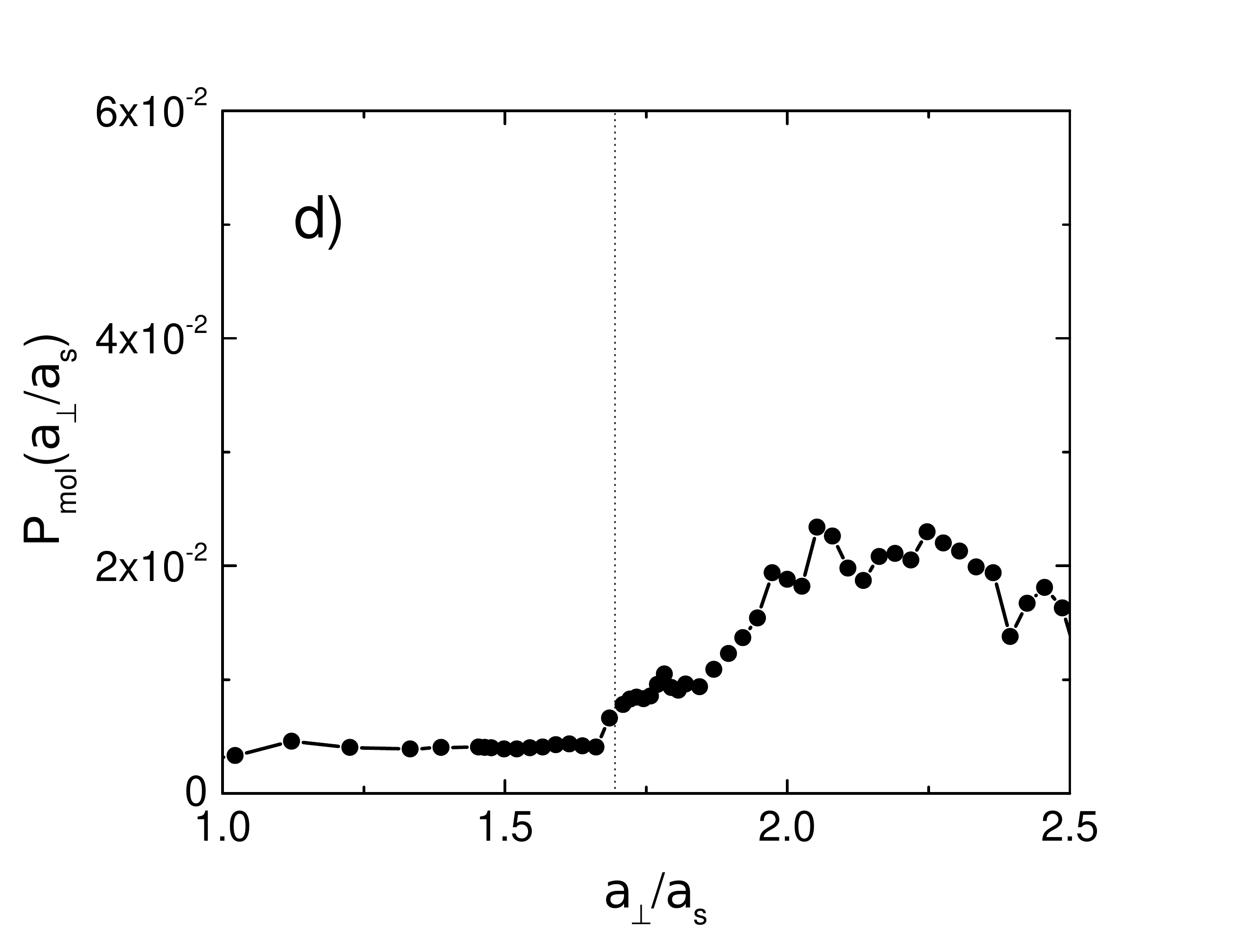}}
\caption{The calculated in the time-independent secular approximation (\ref{eq:sec}) mean ion energy $\langle E^{(out)}_{ik}\rangle$ after Li-Yb$^+$ collision, effective coupling constant $g_{1D}$, transmission coefficient $T$ and the molecular ion formation probability $P_{mol}$ as a function of the ratio $a_{\perp}/a_s$.}
\label{fig:Fig4}
\end{figure}
\begin{figure}
\parbox{0.5\textwidth}{
\centering\includegraphics[scale=0.3]{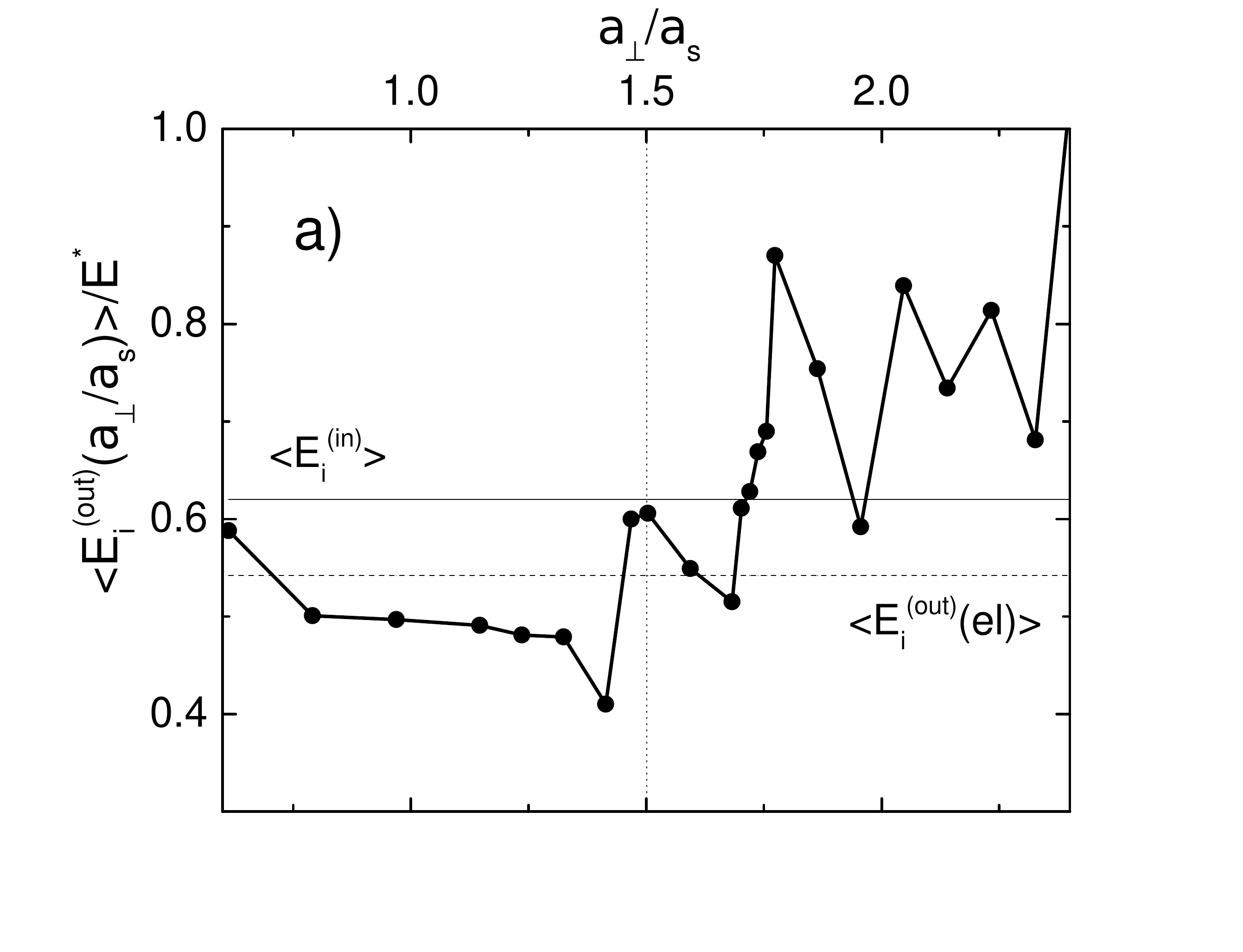}
\vspace{-1.2cm}}
\parbox{0.5\textwidth}{
\centering\includegraphics[scale=0.3]{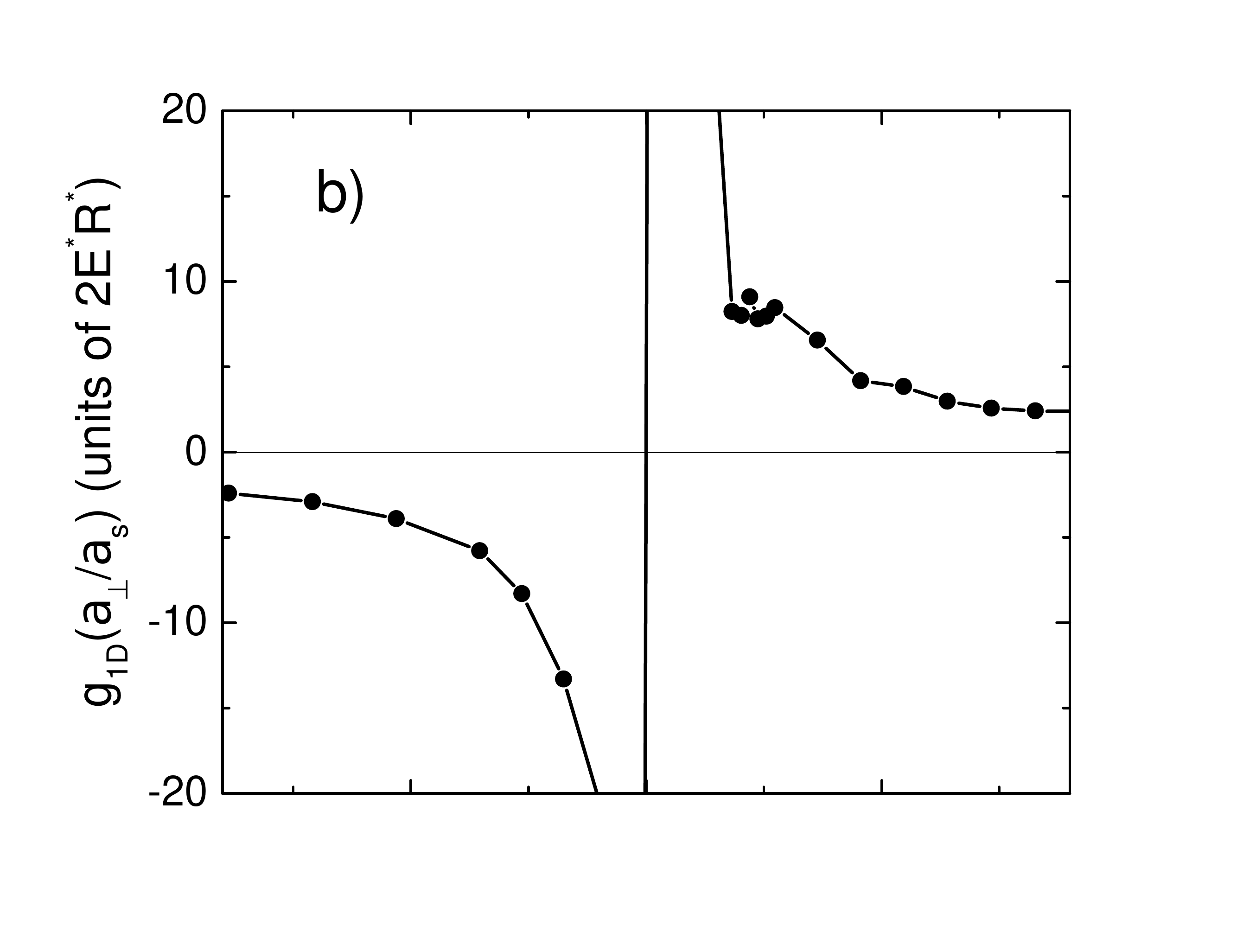}
\vspace{-1.4cm}}
\parbox{0.5\textwidth}{
\centering\includegraphics[scale=0.3]{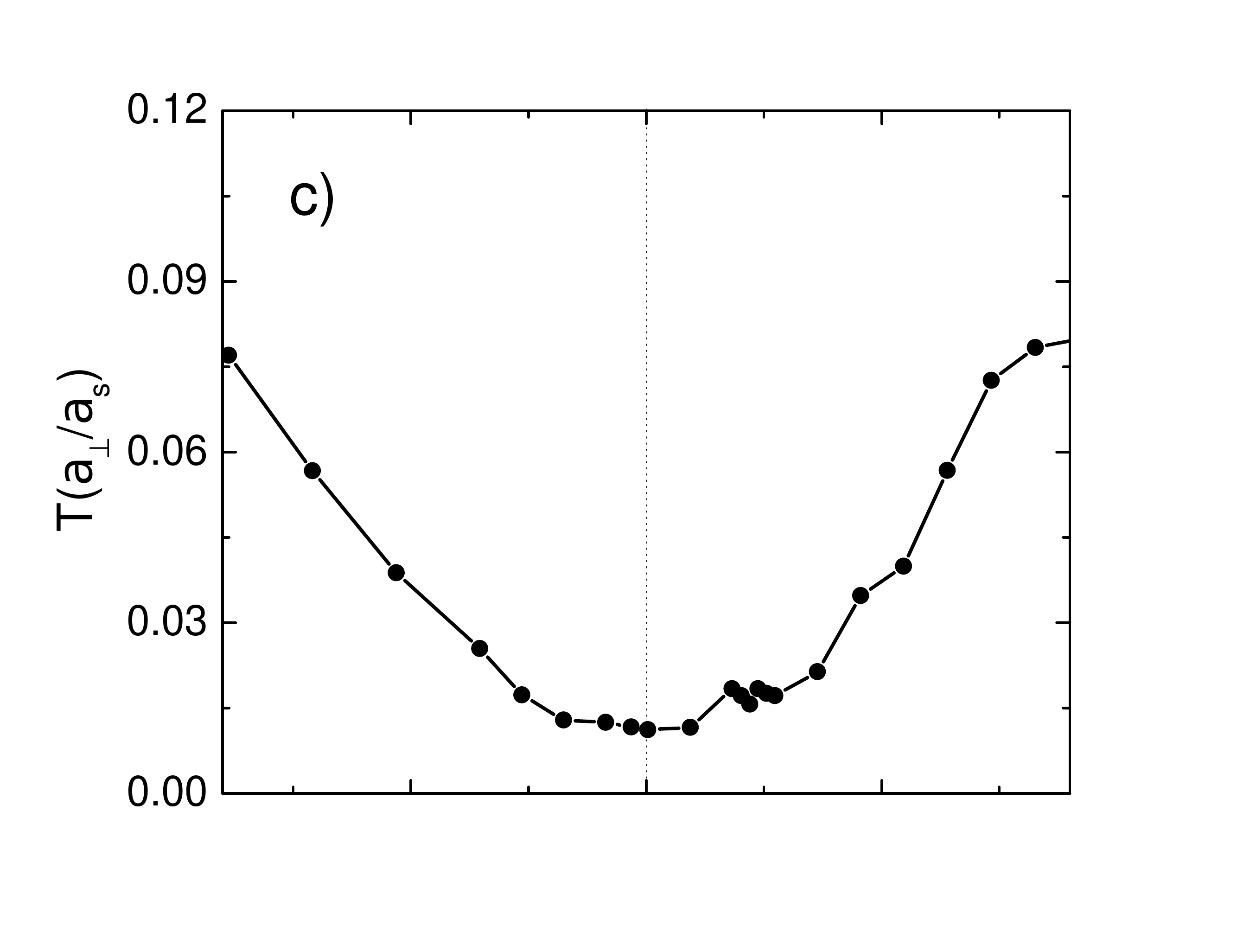}
\vspace{-1.4cm}}
\parbox{0.5\textwidth}{
\centering\includegraphics[scale=0.3]{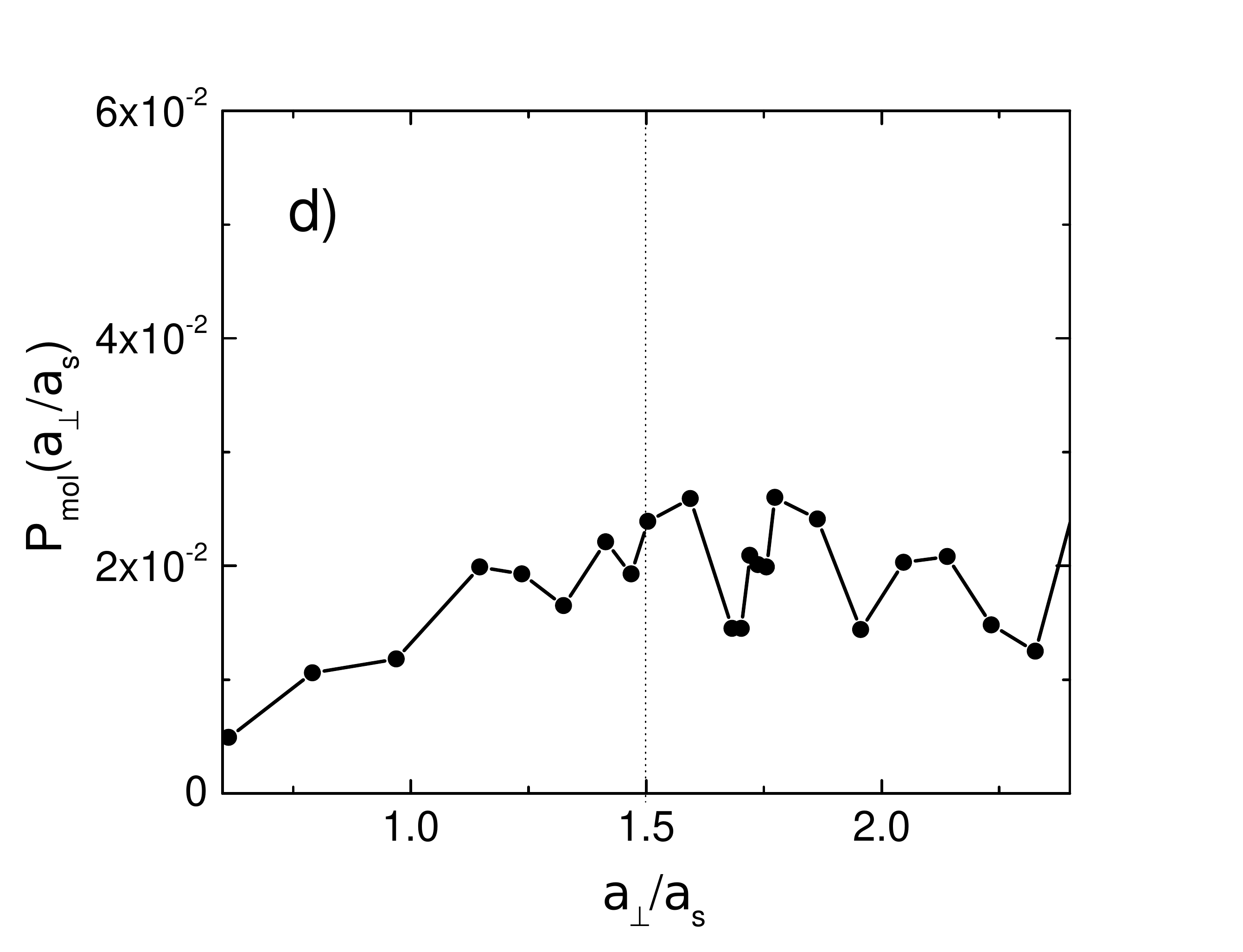}}
\caption{The results of calculations with the potential (\ref{eq:atom-atom}) simulating Li-Yb collision in the time-independent secular approximation (\ref{eq:sec}): the  mean energy $\langle E^{(out)}_{ik}\rangle$ of Yb atom after collision, effective coupling constant $g_{1D}$, transmission coefficient $T$ and the molecule formation probability $P_{mol}$ as a function of the ratio $a_{\perp}/a_s$.}
\label{fig:Fig5}
\end{figure}
The secular equation for the ion trap (\ref{eq:Uion}) gives the following expression~\cite{Melezhik2019}:
\begin{align}
\label{eq:sec}
U_{sec}(\vek{r}_i)=\frac{m_i}{2}[\omega_{xy}^2(x_i^2+y_i^2)+\omega_z^2z_i^2]
\end{align}
with the frequencies $\omega_z=2\pi\times 45$kHz and $\omega_{xy}=2\pi\times 150$ kHz~\cite{Melezhik2019}.
Replacing the time-dependent potential of the interaction of an ion with a Paul trap (\ref{eq:Uion}) into the Hamilton equations (\ref{eq:Hamilton}) by its secular approximation (\ref{eq:sec}), we have calculated the quantities $E_i^{(out)}(a_{\perp}/a_s)$, $g_{1D}(a_{\perp}/a_s)$,$T(a_{\perp}/a_s)$ and $P_{mol}(a_{\perp}/a_s)$ similarly to the calculation performed above. In this case, the initial conditions (\ref{eq:3d}) setting  3D motion of the ion before collisions were used. However, in the case of a stationary secular potential (2), these conditions rigidly fix the mean initial energy of the ion as
$$
\langle E_i^{(in)}\rangle = \frac{1}{2}E_i^{(max)}(t\rightarrow 0)
$$
which is essentially different from the mean energy in the time-dependent Paul trap with the same initial conditions.

The results of the calculations are presented in Fig.\ref{fig:Fig4}. The demonstrated above effect of increasing the efficiency of sympathetic cooling of an ion in the CIR region is also preserved for the secular approximation. Moreover, this region noticeably expands to the left and right of the CIR point, and the observed heating of the ion to the right of the sympathetic cooling region is much less than what we obtained for the time-dependent Paul trap. Indeed, for the RF Paul trap, we observe a sharp increase in the average ion energy to the right of the sympathetic cooling region (see Fig.\ref{fig:Fig3}), almost by an order of magnitude compared to its initial average energy. By replacing the time-dependent Paul trap with a stationary secular approximation, micromotion is eliminated, which leads to the elimination of micromotion-induced heating of the ion. Nevertheless, to the right of the sympathetic cooling region, a slight increase in the ion energy is noticeable. We interpret this growth as the influence of the resonant formation of molecular ions with the transfer of the energy release to the excitation of its center-of-mass. This very narrow (in the absence of the RF part in the Paul trap) resonance state to the right of the "main" resonance at $a_{\perp}/a_s \simeq 1.7$ manifests itself very weakly in the calculated curves $g_{1D}(a_{\perp}/a_s)$ (Fig.\ref{fig:Fig4}b) and $T(a_{\perp}/a_s)$ (Fig.\ref{fig:Fig4}c), but leads to a noticeable increase in the probability $P_{mol}(a_{\perp}/a_s)$ of formation of ionic molecules (Fig.\ref{fig:Fig4}d).

In conclusion of this study, we can summarize that replacing the time-dependent Paul trap with the widely used secular approximation~\cite{LeibfriedRMP03} enhances the effect of increasing the efficiency of sympathetic cooling in the CIR region: this region expands, and the ion cooling depth increases compared to the values achieved in the Paul trap below the limit for the sympathetic cooling following from the model of absolutely elastic head-on collision of mechanical balls.  We explain this effect by the absence in the stationary potential (\ref{eq:sec}) of the time-dependent RF part, which is responsible for the micromotion of ions and, as a consequence, for their possible heating upon collisions with cold atoms, which prevents sympathetic cooling of ions.

\subsection{Sympathetic cooling near CIR in atomic traps}

Finally, we evaluate the effect of the long-range character of the atom-ion interaction on the process of sympathetic cooling. For this purpose we replace the long-range tail in the interparticle interaction (\ref{eq:vai}) by more short-range Van-der-Waals potential
\begin{align}
\label{eq:atom-atom}
V_{aa}(r) =\left \{
\begin{array}{cc} -\frac{(r^2-c^2)}{(r^2+c^2)(r^2+b^2)^2} & ,\,\, r \leq R^* \\ -\frac{(R^{*2}-c^2)}{(R^{*2}+c^2)(R^{*2}+b^2)^2}\frac{R^{*6}}{r^6} & ,\,\, r > R^*
\end{array} \right \}\,.
\end{align}
With this potential we have performed the calculation of the mean kinetic energy of the heavy particle after the collision $\langle E_i^{(out)}(a_{\perp}/a_s)\rangle$, the coupling constant $g_{1D}(a_{\perp}/a_s)$, the transmission coefficient $T(a_{\perp}/a_s)$ and the probability of formation of two-body bound state during the collision $P_{mol}(a_{\perp}/a_s)$. Results of the calculations are presented in Fig.\ref{fig:Fig5}. They were obtained with the potentials (\ref{eq:Uatom}) and (\ref{eq:sec}) describing interactions of the colliding  particles with the trap. Herewith, a trap for a heavy particle was described by the secular approximation from the previous subsection (\ref{eq:sec}). The initial conditions for the heavy particle were specified in the form (\ref{eq:3d}), the same as for the ion in the previous subsection.

The curve $\langle E_i^{(out)}(a_{\perp}/a_s)\rangle$ given in Fig.\ref{fig:Fig5}a qualitatively repeats the behavior of the curve calculated in the previous subsection (Fig.\ref{fig:Fig4}a) in the secular approximation for the ion-trap interaction (\ref{eq:sec}). However, the replacement of the long-range tail in the interparticle interaction (\ref{eq:vai}) by the short-range Van-der-Waals tail (\ref{eq:atom-atom}) leads to a significant narrowing of the region of the minimum of the calculated curve to the right of the CIR and its broadening and deepening to the left of the CIR. We explain this effect by the narrowing of the splitting of CIR into components $(2,i)$ and $(0,i+1)$, which arises in a quasi-1D collision of distinguishable quantum particles, when the long-range potential (\ref{eq:sec}) is replaced by a shorter-range one (\ref{eq:atom-atom}). This splitting is developed in the calculated curves $g_{1D}(a_{\perp}/a_s)$ and $T(a_{\perp}/a_s)$ given in Figs. \ref{fig:Fig5}b and \ref{fig:Fig5}c. Due to the narrowing of the distance between the CIRs $(2,i)$ ($a_{\perp}/a_s$=1.51)and $(0,i+1)$ ($a_{\perp}/a_s\simeq$ 1.72), we observe their practical overlap in Fig.\ref{fig:Fig5}d, which shows the calculated probability $P_{mol}$ of the formation of a bound two-particle system in a collision. Moreover, since the splitting between the components $(2,i)$ and $(0,i+1)$ of the CIR is very narrow, the resonant amplification of $P_{mol}$ in the point of CIR $(0,i+1)$ also covers the region of the resonance $(2,i)$. Note also that the replacement of the long-range tail in the interparticle interaction by the short-range Van-der-Waals potential (\ref{eq:atom-atom}) leads to a significant shift of the CIR resonances to the left: the left resonance $(2,i)$ shifts from point $a_{\perp}/a_S=1.70$ (Fig.\ref{fig:Fig4}b)to point $a_{\perp}/a_S=1.51$ (Fig.\ref{fig:Fig5}b). 

In conclusion, the calculation performed shows that  the replacement of the long-range tail in the atom-ion interaction (\ref{eq:vai}) by the short-range Van-der-Waals potential (\ref{eq:atom-atom}), which simulates the atom-atom interaction, does not fundamentally change the picture of enhanced sympathetic cooling of a heavy hot particle near CIR. Moreover, in this case we get a deeper minimum for the curve $\langle E_i^{(out)}(a_{\perp}/a_s)\rangle$ to the left of CIR, which is much deeper than the limit following from the model of absolutely elastic central collision of two balls.

The above consideration can be regarded as a rough model of the quasi-1D Li-Yb scattering in an optical trap in the vicinity of the atomic CIR, which demonstrates that the effect of enhancing the sympathetic cooling of the $^+$Yb-ion in confined collisions with Li-atoms in the resonant region also remains in the confined collision of atomic Yb with cold Li-atoms. An obvious improvement of the model is the use of more realistic parameters $r^* \simeq 0.03R^*$ for matching the Van-der-Waals tail with the inner part of the potential (\ref{eq:atom-atom}) instead of $r^*=R^*$ we use here and $C_6$-coefficient, as well as more realistic parameters $\omega_{x,y}$ and $\omega_z$ in the interaction of Yb-atom with the optical trap (\ref{eq:sec}). Such a modification, however, requires a significant improvement in the convergence and accuracy of the computational scheme we use here.

\section{CONCLUSION}

We have investigated the effect of sympathetic cooling around CIRs in atom-ion and atom-atom collisions with qusiclassical-quantum approach using the Li-Yb$^+$ and Li-Yb confined systems as an example. In this approach, the Schr\"odinger equation for a cold light atom is integrated simultaneously with the classical Hamilton equations for a hotter heavy ion (atom) during collision. We have shown in the framework of this model that the region near the atom-ion CIR is the most promising for the sympathetic cooling of an ion by cold atoms in a hybrid atom-ion trap due to suppression of the micromotion-induced heating.
The origin of this suppression is the ``fermionization'' effect of the relative distribution of the atom-ion probability density near CIR, where the atom-ion pair behaves as a pair of noninteracting identical fermions, which partially compensates the long-range character of the atom-ion interaction and, as a consequence, the possibility to enhance the ion micromotion due to collisions.

Moreover, it was also demonstrated that in the absence of micromotion of a heated particle, as in the case of the secular approximation for an ion in a hybrid atom-ion trap or in the case of atoms in an optical trap, the efficiency of its sympathetic cooling in the CIR region increases.

Thus, based on the performed analysis we propose a new way for sympathetic cooling of ions in an electromagnetic Paul trap: to use for this purpose cold buffer atoms in the region of atom-ion confinement-induced resonance (CIR). It was also shown that it is possible to improve the efficiency of sympathetic cooling in atomic traps by using atomic CIRs.

In conclusion, atom-ion CIRs have not been experimentally discovered yet. Therefore, it seems to us that it is easier to varify experimentally the predicted mechanism of sympathetic cooling in atomic systems, where magnetic Feshbach resonances are successfully used to tune atomic CIRs. Nevertheless, the implementation and tuning of the effective atomic-ion interaction on the atom-ion CIRs may turn out to be a simpler experimental problem than in pure atomic confined systems that does not require the use of the Feshbach magnetic resonance technique. Indeed, the nonresonant s-wave atomic scattering lengths $a_s$ in free space are much smaller than the transverse dimensions of the existing atomic optical traps $a_{\perp}$, that is, in the nonresonant case  $a_{\perp} \gg \mid a_s \mid $. Therefore, the realization of the resonance condition $a_{\perp} = 1.46 a_s$ for atomic systems required the use of the technique of magnetic Feshbach resonances for a sharp increase in the value of $a_s$. In atom-ion systems, the scattering lengths, even in the nonresonant case, significantly exceed the characteristic nonresonant atomic scattering lengths due to the long-range nature of atomic-ion interactions compared to atomic interactions, which leads to the fulfillment of the relation $ a_{\perp} \sim |a_s| $ for atom-ion systems already in the nonresonant case. Therefore, the necessary fine tuning of the atomic-ion scattering length $ a_s $ to satisfy the resonance condition $a_{\perp}= 1.46a_s$ can here be replaced by just a slight variation of the atomic trap width $a_{\perp} $ without using the Feshbach resonance technique for enhancement of $a_s$. In this context, the task of experimental detection of the atom-ion CIR and its use for sympathetic cooling of ions in a hybrid atom-ion trap seems to us  quite feasible and relevant.

Note also  that a fully quantum consideration of the problem would be very useful, especially in the case of comparable masses of an ion and an atom, when quantum effects become significant. However, its realization is rather challenging technical problem but realistic one.

\section{ACKNOWLEDGEMENTS}
The author thanks P.Schmelcher, A. Negretti, Z. Idziaszek, and O. Prudnikov for fruitful discussions. The work was supported by the Russian Foundation for Basic Research, Grants No. 18-02-00673.

\section{Appendix}
The coupled system of quantum (\ref{eq:Schr}) and classical (\ref{eq:Hamilton}) equations describes
collisional dynamics of a Li-Yb$^+$ pair confined in a hybrid trap with three absolutely
different time-scales $2\pi/\Omega_{rf} \ll 2\pi/\omega_i \ll t_{\perp} = 2\pi/\omega_{\perp}$ defined
by Paul trap frequencies of  ($\Omega_{rf} = 2\pi \times$ 2 MHz and $\omega_i = 2\pi \times 63$ kHz) and atomic waveguide ($\omega_{\perp} \simeq 2\pi \times 10$ kHz). These three time scales impose strict requirements
on the computational scheme. The scheme must be stable enough
long time-interval (time of atom-ion collision) $\sim 10\times t_{\perp} = 10 \times  2\pi/\omega_{\perp}$ and, on
the other hand, it must accurately handle fast oscillations defined by the
frequency $\Omega_{rf}$ of the RF-field, as well as the resonant behavior of the the atom-ion interaction potential $V_{ai}(\mid \vek{r}_a-\vek{r}_i(t)\mid)$ (\ref{eq:vai}) on collision.

To integrate the coupled system of equations of motion (\ref{eq:Schr}) and (\ref{eq:Hamilton}), we applied the splitting-up method
with a 2D discrete-variable representation (DVR)~\cite{dvr1,dvr2,dvr3}.
For an accurate inclusion of the atom-ion interaction potential (\ref{eq:vai}) in the numerical integration procedure of the Schr\"odinger equation (\ref{eq:Schr}) at the moment of the resonant atom-ion collision, a tailored splitting-up procedure in the 2D-DVR representation was developed in ~\cite{Melezhik2019}.

Simultaneously to the forward in time propagation
$t_n\rightarrow t_{n+1}=t_n+\Delta t$ of the atom wave packet $\psi(\vek{r}_a,t_{n})\rightarrow \psi(\vek{r}_a,t_{n+1})$ when integrating the time-dependent Schr\"odinger equation (\ref{eq:Schr}), we integrate
the Hamilton equations of motion ~(\ref{eq:Hamilton}), which involve three different scales of frequencies, namely $\Omega_{rf}$, $\omega_i$, as well as $\omega_{\perp}$ in the quantum mechanical average  $\langle\psi(\vek{r}_a,t;\vek{r}_i)\vert V_{ai}(|\hat{\vek{r}}_a-\vek{r}_i(t)|)\vert\psi(\vek{r}_a,t;\vek{r}_i)\rangle$. To this end, we have adapted the second-order St\"ormer-Verlet method~\cite{Verlet} to our problem

\begin{equation}
\vek{p}_i^{(n+1/2)} = \vek{p}_i^{(n)} - \frac{\Delta t}{2}\frac{\partial}{\partial \vek{r}_i}H_i(\vek{p}_i^{(n+1/2)},\vek{r}_i^{(n)})\,,\nonumber
\end{equation}
\begin{align}
\vek{r}_i^{(n+1)} = \vek{r}_i^{(n)} + \frac{\Delta t}{2}\left\{\frac{\partial}{\partial \vek{r}_i}H_i(\vek{p}_i^{(n+1/2)},\vek{r}_i^{(n)})
\right.\nonumber \\
\left.+\frac{\partial}{\partial \vek{r}_i}H_i(\vek{p}_i^{(n+1/2)},\vek{r}_i^{(n+1)})\right\}\,,\nonumber
\end{align}
\begin{equation}
\vek{p}_i^{(n+1)} = \vek{p}_i^{(n+1/2)} - \frac{\Delta t}{2}\frac{\partial}{\partial \vek{r}_i}H_i(\vek{p}_i^{(n+1/2)},\vek{r}_i^{(n+1)})\,.
\end{equation}
 Here,$\vek{p}_i^{(n)}=\vek{p}_i\left(t_n\right)$, $\vek{p}_i^{(n+1/2)}=\vek{p}_i\left(t_n+\frac{\Delta t}{2}\right)$, and $\vek{p}_i^{(n+1)}=\vek{p}_i\left(t_n+\Delta t\right)$ and the same definition for $\vek{r}_i^{(n)}$.

The convergent results were obtained with a time step of the order of $\Delta t = t_{\perp}/6000$.


\begin{thebibliography}{40}
\bibitem{Cote} R.~C\'ot\`e, V. Kharchenko, and M. D. Lukin, Mesoscopic molecular ions in Bose-Einstein condensates. Phys. Rev. Lett. {\bf 89}, 093001 (2002).
\bibitem{Shuher} J. M. Schurer, A. Negretti, and P. Schmelcher, Unraveling the structure of ultracold mesoscopic collinear molecular ions. Phys. Rev. Lett. {\bf 119}, 1 (2017).
\bibitem{Bissbort} U. Bissbort, D. Cocks, A. Negretti, Z. Idziaszek, T. Calarco, F. Schmidt-Kaler, W. Hofstetter, and R. Gerritsma, Emulating solid-state physics with a hybrid system of ultracold ions and atoms. Phys. Rev. Lett. {\bf 111}, 080501 (2013).
\bibitem{Idziaszek} Z. Idziaszek, T. Calarco, P. S. Julienne, and A. Simoni, Quantum theory of ultracold atom-ion collisions. Phys. Rev. A {\bf 79}, 010702 (2009).
\bibitem{Moszynski} M. Tomza, C. P. Koch, and R. Moszynski, Cold interactions between an Yb$^+$ ion and a Li atom: prospects for sympathetic cooling, radiative association, and Feshbach resonances. Phys. Rev. A{\bf 91}, (2020).
\bibitem{Doerk} H. Doerk, Z. Idziaszek, and T. Calarco, Atom-ion quantum gate. Phys. Rev. A{\bf 81}, 012708 (2010).
\bibitem{Secker} T. Secker, R. Gerritsma, A. W. Glaetzle, and A. Negretti, Controlled long-range interactions between Rydberg atoms and ions. Phys. Rev. A{\bf 94}, 013420 (2016).
\bibitem{Tomza} M. Tomza, K. Jachumski, R. Gerritsma, A. Negretti, T. Calarco, Z. Idziaszek, and P. S. Julienne, Cold hybrid atom-ion systems. Rev. Mod. Phys. {\bf 91}, 03500 (2019).
\bibitem{Grier} A. T. Grier, M. Cetina, F. Oru\`cevi\'c, and V. Vuleti\'c, Observation of Cold Collisions between Trapped Ions and Trapped Atoms. Phys. Rev. Lett. {\bf 102}, 223201 (2009).
\bibitem{Meir} Z. Meir, T. Sikorsky, R. Ben-shlomi, N. Akerman, Y. Dallal, and R. Ozeri, Dynamics of a ground-state cooled ion colliding with ultracold atoms. Phys. Rev. Lett. {\bf 117}, 243401 (2016).
\bibitem{Furst} H. A. F\"urst, N. V. Ewald, T. Secker, J. Joger, T. Feldker, and R. Gerritsma, Prospects of reaching the quantum regime in Li-Yb$^+$ mixtures. J. of Phys. B{\bf 51}, 195001 (2018).
\bibitem{Vuletic} M. Cetina, A.T. Grier, and V. Vuleti\'c, Micromotion-induced limit to atom-ion sympathetic cooling in Paul traps, Phys. Rev. Lett. {\bf 109}, 253201 (2012).
\bibitem{Zipkes} C. Zipkes, S. Palzer, C. Sias, and M. K\"ohl, A trapped single ion inside a Bose-Einstein condensate, Nature {\bf 464}, 388 (2010).
\bibitem{Ravi} K. Ravi, S. Lee, A. Sharma, G. Werth, and S. Rangwala, Cooling and stabilization by collisions in a mixed ion-atom system. Nature Communications {\bf 3}, 1126 (2012).
\bibitem{Harster} A. H\"arter and J. H. Denschlag, Cold atom-ion experiments in hybrid traps. Contemporary Physics {\bf 55}, 33 (2014).
\bibitem{Haze} S. Haze, M. Sasakawa, R. Saito, R. Nakai, and T. Mukaiyama, Cooling dynamics of a single trapped ion via elastic collisions with small-mass atoms. Phys. Rev. Lett. {\bf 120}, 043401 (2018).
\bibitem{Smith} I. Sivarajah, D. S. Goodman, J. E.Wells, F. A. Narducci, and W. W. Smith, Evidence of sympathetic cooling of Na$^+$ ions by a Na magneto-optical trap in a hybrid trap. Phys. Rev. A{\bf 86}, 063419 (2012).
\bibitem{Feldker} T. Feldker, H. F\"urst, H. Hirzler, N. V. Ewald, M. Mazzanti, D. Wiater, M. Tomza, and R. Gerritsma, Buffer gas cooling of a trapped ion to the quantum regime. Nature Physics {\bf 16}, 413 (2019).
\bibitem{Kleinbach} K. S. Kleinbach, F. Engel, T. Dieterle, R. L\"ow, T. Pfau, and F. Meinert, Ionic impurity in a Bose-Einstein condensate at submicrokelvin temperatures, Phys. Rev. Lett. {\bf 120}, 193401 (2018).
\bibitem{Prudnikov} O.N.~Prudnikov, S.V.~Chepurov, A.A.~Lugovoy, K.M.~Rumynin, S.N.~Kuznetsov, A.V.~Taichenachev, V.I.~Yudin, S.N.~Bagayev, Laser cooling of
171Yb+ ions in a frequency-modulated field. Quantum Electronics {\bf 47}, 806 (2017).
\bibitem{MelNegr} V.S.~Melezhik and A.~Negretti, Confinement-induced resonances in ultracold atom-ion systems. Phys. Rev. A{\bf 94}, 022704 (2016).
\bibitem{Melezhik2019} V. Melezhik, Z. Idziaszek, and A. Negretti, Impact of ion motion on atom-ion confinement-induced resonances in hybrid traps. Phys. Rev. A{\bf 100}, 063406 (2019).
\bibitem{Olshanii} M.~Olshanii, Atomic scattering in the presence of an external confinement and a gas of impenetrable bosons. Phys. Rev. Lett. {\bf 81}, 938 (1998).
\bibitem{Haller2010} E.~Haller, M.J.~Mark, R.~Hart, J.G.~Danzl, L.~Reichs\"ollner, V.~Melezhik, P.~Schmelcher,
and H.C. N\"agerl, Confinement-Induced Resonances in Low-Dimensional Quantum Systems. Phys. Rev. Lett. {\bf 104}, 153203 (2010)
\bibitem{Bergeman} T.~Bergeman, M.G.~Moore, and M. Olshanii, Atom-atom scattering under cylindrical harmonic confinement: numerical
and analytic studies of the confinement induced resonance. Phys. Rev. Lett. {\bf 91}, 163201 (2003).
\bibitem{Moore} M.G.~Moore, T.~Bergeman, and M.~Olshanii, J. Phys. IV France {\bf 116}, 69 (2004).
\bibitem{Kim} J.~Kim, V.~Melezhik, and P.~Schmelcher, Suppression of quantum scattering in strongly confined systems. Phys. Rev. Lett. {\bf 97}, 193203 (2006).
\bibitem{Naidon} P.~Naidon, E.~Tiesinga, W.~Mitchell, and P.~Julienne, Effective-range description of a Bose gas under strong one- or two-dimensional confinement. New J. Phys. {\bf 9}, 19 (2007).
\bibitem{Saeidian} S.~Saeidian, V.S.~Melezhik, and P.~Schmelcher, Multichannel atomic scattering and confinement-induced resonances in waveguides. Phys. Rev. A{\bf 77}, 042721 (2008).
\bibitem{Mel2011} V.S.~Melezhik and P.~Schmelcher, Multichannel effects near confinement-induced resonances in harmonic waveguides. Phys. Rev. A{\bf 84}, 042712 (2011).
\bibitem{Giannakeas} P.~Giannakeas, V.~Melezhik, and P.~Schmelcher, Dipolar confinement-induced resonances of ultracold gases in waveguides. Phys. Rev. Lett. {\bf 111}, 183201 (2013).
\bibitem{Gunter} K.~G\"unter, T.~St\"oferle, H.~Moritz, M.~K\"ohl, and T.~Esslinger, p-Wave interactions in low-dimensional fermionic gases. Phys. Rev. Lett. {\bf 95}, 230401 (2005).
\bibitem{Frolich} B.~Fr\"ohlich, M.~Feld, E.~Vogt, M.~Koschorreck, W.~Zwerger, and M.~K\"ohl, Radio-frequency spectroscopy of a strongly interacting two-dimensional Fermi gas. Phys. Rev. Lett. {\bf 106}, 105301 (2011).
\bibitem{Kinoshita} T.~Kinoshita, T.~Wenger, and D.S.~Weiss, Observation of a one-dimensional Tonks-Girardeau gas. Science {\bf 305}, 1125 (2004).
\bibitem{Peredes} B.~Paredes, A.~Widera, V.~Murg, O.~Mandel, S.~F\"olling, I.~Cirac, G.V.~Shlyapnikov, T.W.~H\"ansch, and I.~Bloch, Tonks–Girardeau gas of ultracold atoms in an optical lattice. Nature (London) {\bf 429}, 277 (2004).
\bibitem{Haller2009} E.~Haller, M.~Gustavsson, M.J.~Mark, J.G.~Danzl, R.~Hart, G.~Pupillo, and H.-C.~N\"agerl, Realization of an excited, strongly correlated quantum gas phase science {\bf 325}, 1224 (2009).
\bibitem{Selim} G.~Z\"urn, F.~Serwane, T.~Lompe, A.N.~Wenz, M.G.~Ries, J.E.~Bohn, and S.~Jochim, Fermionization of Two Distinguishable Fermions. Phys. Rev. Lett. {\bf 108}, 075303 (2012).
\bibitem{Girardeau} M.~Girardeau, Relationship between Systems of Impenetrable Bosons and Fermions in One Dimension. J. Math. Phys. {\bf 1}, 516 (1960).
\bibitem{MelSchm} V.S.~Melezhik and P.~Schmelcher, Quantum energy flow in atomic ions moving in magnetic fields. Phys. Rev. Lett. {\bf 84}, 1870 (2000).
\bibitem{Melezhik2001} V.S.~Melezhik, Recent progress in treatment of sticking and stripping with time-dependent approach. Hypefine Int. {\bf 138}, 351 (2001).
\bibitem{MelezhikCohen} V.S.~Melezhik, J.S.~Cohen, and C.Y.~Hu, Stripping and excitation in collisions between p and He$^+$(n$\leq$3) calculated by a quantum time-dependent approach with semiclassical trajectories. Phys. Rev. A{\bf 69}, 032709 (2004).
\bibitem{MelSev} V.S.~Melezhik and L.A.~Sevastianov, Quantum-semiclassical calculation of transition probabilities in proton collisions with helium ions. Analytical and Computational Methods in Probability Theory,  LNCS {\bf 10684}, 449 (2017).
\bibitem{JogerPRA17} J. Joger, H. F\"urst, N. Ewald, T. Feldker, M. Tomza, and R. Gerritsma, Observation of collisions between cold Li atoms and Yb$^+$ ions. Phys. Rev. A{\bf 96}, 030703(R) (2017)
\bibitem{FuertsPRA18} H. F\"urst, T. Feldker, N. V. Ewald, J. Joger, M. Tomza, and R. Gerritsma, Dynamics of a single ion-spin impurity in a spin-polarized atomic bath. Phys. Rev. A{\bf 98}, 012713 (2018).
\bibitem{Mel2019} V.S.~Melezhik, Efficient computational scheme for ion dynamics in RF-field of Paul trap. Discrete and Continuous Models and Applied Computational Science {\bf 27 (4)}, 378 (2019); DOI: 10.22363/2658-4670-2019-27-4-378-385.
\bibitem{LeibfriedRMP03} D. Leibfried, R. Blatt, C. Monroe, and D. Wineland, Quantum dynamics of single trapped ions. Rev. Mod. Phys. {\bf 75}, 281 (2003).
\bibitem{Krych2015} M.~Krych and Z.~Idziaszek, Description of ion motion in a Paul trap immersed in a cold atomic gas. Phys. Rev. A{\bf 91}, 023430 (2015).
\bibitem{Chin2010} C.~Chin, R.~Grimm, P.S.~Julienne, and E.~Tiesinga, Feshbach resonances in ultracold gases. Rev. Mod. Phys. {\bf 82}, 1225
\bibitem{Thorwart} V.~Peano, M.~Thorwart, C.~Mora, and R.~Egger, Confinement-induced resonances for a two-component gas in arbitrary quasi-one-dimensional traps. New J. Phys. {\bf 7}, 192 (2005).
\bibitem{Melezhik2009} V.S.~Melezhik and P.~Schmelcher, Quantum dynamics of resonant molecule formation in waveguides. New J. Phys. {\bf 11}, 073031 (2009).
\bibitem{Sala} S.~Sala, G.~Z\"urn, T.~Lompe, A.N.~Wenz, S.~Murmann, F.~Serwane, S.~Jochim, and A.~Saenz, Coherent molecule formation in anharmonic potentials near confinement-induced resonances. Phys. Rev. Lett. {\bf 110}, 203202 (2013).
\bibitem{dvr1} V.S.~Melezhik, Nondirect product discrete variable representation in multidimensional quantum problems, in {\it Numerical Analysis and Applied Mathematics ICNAAM2012: International Conference of Numerical Analysis
and Applied Mathematics}, edited by T. E. Simos, G. Psihoyios, C. Tsitouras, and Z. Anastassi, AIP Conf. Proc. No. 1479 (AIP, New York, 2012), p.1200.
\bibitem{dvr2} V.S.~Melezhik, A computational method for quantum dynamics of a three-dimensional atom in strong fields, in
{\it Atoms and Molecules in Strong External Fields}, edited by P. Schmelcher and W. Schweizer (Plenum, New York, 1998), p.89.
\bibitem{dvr3} V.S.~Melezhik, Polarization of harmonics generated from a hydrogen atom in a strong laser field. Phys. Lett. A{\bf 230}, 203 (1997).
\bibitem{Verlet} E.~Hairer, Ch.~Lubich, and G.~Wanner. {\it Geometric Numerical integration. Structure-Preserving Algorithms for Ordinary Differential Equations} (Springer-Verlag, Berlin Heidelberg, 2006).Ch.I.
\end{thebibliography}
\end{document}